\documentclass[a4paper,12pt]{article}
\usepackage[paper=letterpaper,margin=1in]{geometry}
\usepackage{graphicx}
\usepackage{hyperref}
\hypersetup{
    colorlinks,
    citecolor=red,
    filecolor=black,
    linkcolor=blue,
    urlcolor=black
}

\usepackage{amsmath,amssymb,amsfonts,epsfig,cite,setspace,bigstrut, verbatim}
\pdfoutput=1

\makeatother
\date{}
\numberwithin{equation}{section}

\begin{document}

\title{\textbf{Dessins d'Enfants in $\mathcal{N}=2$ Generalised Quiver Theories}}

\author{Yang-Hui He\textsuperscript{1} and
James Read\textsuperscript{2}
}
\begin{onehalfspace}
\maketitle
\begin{center}
\textsuperscript{1}\emph{Department of Mathematics, City University,
London,}\\
\emph{Northampton Square, London EC1V 0HB, UK;}\\
\emph{School of Physics, NanKai University, Tianjin, 300071, P.R.
China;}\\
\emph{Merton College, University of Oxford, OX1 4JD, UK}\\
\emph{hey@maths.ox.ac.uk}\\
~\\
\textsuperscript{2}\emph{Merton College, University of Oxford, OX1 4JD, UK}\\
\emph{james.read@merton.ox.ac.uk}\\
\par\end{center}

\medskip{}

\begin{abstract}
We study Grothendieck's \emph{dessins d'enfants} in the context of the $\mathcal{N}=2$ supersymmetric gauge theories in  $\left(3+1\right)$ dimensions with product $SU\left(2\right)$ gauge groups which have recently been considered by Gaiotto \emph{et al}.  We identify the precise context in which dessins arise in these theories:~they are the so-called ribbon graphs of such theories at certain isolated points in the moduli space. With this point in mind, we highlight connections to other work on trivalent dessins, gauge theories, and the modular group.
\end{abstract}
\end{onehalfspace}

\pagebreak{}

\tableofcontents{}

\section{Introduction}

In the mid-1990s, the pioneering work of Seiberg and Witten \cite{SeibergWitten} led to
a revolution in the study of $\mathcal{N}=2$ supersymmetric gauge
theories in  $\left(3+1\right)$ dimensions. Their work, which has come to be known as
 {Seiberg-Witten theory}, deals with the construction of the non-perturbative
dynamics of $\mathcal{N}=2$ theories in the limit of low energy
and momenta. The jewel in the crown
of Seiberg-Witten theory is the \textbf{Seiberg-Witten curve}:~a (hyper)elliptic curve,
the periods of which completely specify the spectra, coupling, and low-energy
effective Lagrangian, as well as non-perturbative information of the
gauge theory.

In this paper, we undertake a study of a class of $\mathcal{N}=2$
theories known as \textbf{Gaiotto theories}. These theories have interesting
duality behaviour amongst themselves under S-duality and possess a systematic
string-theoretic construction for their Seiberg-Witten curves \cite{HeJohn}:~take
a Riemann surface $\mathcal{C}$ of genus $g$ with $n$ punctures,
dubbed the \textbf{Gaiotto curve}\emph{,} and wrap $N$ coincident
parallel M5-branes over it. The world-volume theory on the M5-brane is one with a product
$SU\left(N\right)$ group, the Seiberg-Witten curve for which is an
$N$-fold cover over $\mathcal{C}$.

Let us henceforth take $N=2$. Each such Gaiotto theory has a product
$SU\left(2\right)^{3g-3+n}$ gauge group \cite{Hanany}, and all can
be encoded into a \textbf{skeleton diagram}, also known as a \textbf{generalised
quiver diagram} \cite{Hanany}. This is a trivalent graph with internal
edges corresponding to $SU\left(2\right)$ groups and external legs
corresponding to flavours. Hence we have a graph with $g$ closed
circuits, $3g-3+n$ edges, $2g-2+n$ nodes, and $n$ external lines.
The diagram constitutes the spine of the am{\oe}ba projection of the Gaiotto
curve%
\cite{amb}, and hence captures the genus and number of punctures on
$\mathcal{C}$. Each skeleton diagram determines a unique $\left(3+1\right)$
dimensional $\mathcal{N}=2$ gauge theory \cite{HeJohn}.



To illustrate, the skeleton diagram for the Gaiotto
theory with a single $SU\left(2\right)$ factor and $N_{f}=4$ flavours (which, in virtue of its simplicity, we shall use as a running example)
is shown in Figure 1(a). Note that in this paper, $N_f$ always denotes the number of $SU\left(2\right)$ factors in the flavour symmetry group, rather than the number of so-called \emph{fundamental flavours} which appear in the linear quivers drawn in e.g.~\cite{Gaiotto, GMN}. The corresponding \emph{BPS quiver}, to be discussed
below, is drawn in Figure 1(b).

\begin{figure}

\begin{center}
\begin{minipage}[t]{0.47\textwidth}%
\begin{center}
\includegraphics{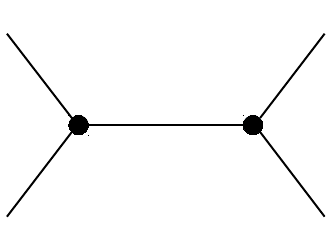}
\par
(a)\end{center}

\end{minipage}\qquad{}%
\begin{minipage}[t]{0.47\textwidth}%
\begin{center}
\includegraphics{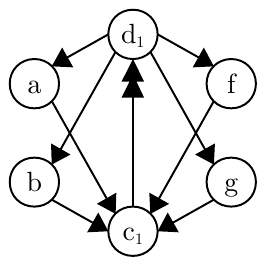}
\par
(b)\end{center}

\end{minipage}
\par\end{center}

\sf{\caption{(a) Skeleton diagram and (b) BPS quiver for the Gaiotto theory with a single $SU\left(2\right)$ factor
and $N_{f}=4$.}}
\end{figure}

In a parallel vein, the Seiberg-Witten curves for $SU\left(2\right)$ Gaiotto theories
can be written in the form $y^{2}=\phi\left(x\right)$, where $q=\phi\left(x\right)\mathrm{d}x^{2}$
is a \textbf{quadratic differential} on $\mathcal{C}$ with only second
order poles \cite{Gaiotto,MainVafa}. The functional form of $q$ is specified
by the topology $\left\langle g,n\right\rangle $ of the skeleton;
varying individual parameters in $q$ amounts to changing the
point in the moduli space of the theory under consideration.
With this in mind, an obvious question now arises: given a specific skeleton diagram,
 how do we extract the relevant Seiberg-Witten curve for that Gaiotto theory, as well as other
 information about the theory, such as that relating to its BPS spectrum?

Though the answer
to this question turns out to be simple, consideration as to
how to respond leads us to an intricate
web of recently-discovered structures important in the study of $SU\left(2\right)$ Gaiotto theories.
In addition to the structures introduced above, this web includes the so-called \textbf{BPS quivers}
which arise in the BPS spectroscopy of the theory, in addition to several important
graphs drawn on $\mathcal{C}$. Many of the structures in
this web have already been carefully elaborated in the recent work on BPS quivers in the context of 
Gaiotto theories (see \cite{MainVafa, CecottiVafa, FirstVafa}), and in other work on Gaiotto theories (\cite{Gaiotto, Hanany, HeJohn}, etc.). By presenting this web, we are able to precisely identify where \textbf{dessins d'enfants}, i.e.~bipartite graphs drawn on
Riemann surfaces, arise in the context of these Gaiotto theories, thereby in turn allowing us to connect the study of $SU\left(2\right)$ Gaiotto theories to previous work on dessins in the context of $\mathcal{N}=2$ theories \cite{FirstCachazo, CachazoDessins, HeJohn}.

Specifically, it turns out that dessins arise in the context of these theories as so-called \textbf{ribbon graphs} on $\mathcal{C}$,
at isolated points in the Coulomb branch of the moduli space where the quadratic differential on $\mathcal{C}$ satisfies
the definition of a so-called \textbf{Strebel differential}, and further where $n$ additional real numbers associated to the punctures are tuned such that the edges of the ribbon graphs have equal lengths \cite{Mulase}.
 Recognising this yields many results. First, by 
\textbf{Belyi's theorem}, we find that Gaiotto curves at such points have the structure of algebraic 
curves defined over $\mathbb{\overline{Q}}$. Next, results from \cite{Mulase} yield an efficient means of computing
the explicit  Strebel differentials, and therefore Seiberg-Witten curves, at these points in the moduli space, via
the dessin's associated \textbf{Belyi map}. In addition, we are led to connections with the dessins in \cite{YMR}, which
correspond to certain subgroups of the {modular group}, and to conjectures on further connections
to the work of \cite{FirstCachazo, CachazoDessins} on dessins and $\mathcal{N}=2$, $U\left(N\right)$ gauge
theories. 

With an understanding of all these structures and connections, in particular of the role of dessins in the study of 
these Gaiotto theories, we proceed in the final part of this paper to study an alternative proposal for 
the role of dessins in $SU\left(2\right)$ Gaiotto theories. Specifically,
it has recently been suggested in \cite{HeJohn} that the skeleton diagrams
themselves can be interpreted as dessins d'enfants, from which one can extract
the relevant Seiberg-Witten curve by manipulating the associated {Belyi map}. We shall
show that such a suggestion needs to be modified in general, and shall provide a number of examples to 
highlight this.

The structure of this paper is as follows. In \S2, we introduce all the key structures which
arise in the study of $SU\left(2\right)$ Gaiotto theories. In \S3, we elaborate the connections 
between these structures, providing an extended discussion of the role of dessins d'enfants in the
context of these theories. Finally, in \S4 we evaluate the above-mentioned alternative proposals for how dessins arise in the context of these theories.


\section{Dramatis Person{\ae}}

In this section, we present a pedagogical summary of all the important mathematical structures
which arise in the study of $SU\left(2\right)$ Gaiotto theories; the web of interrelations
between these structures shall then be elaborated in the following section. In \S2.1, we
present some more technical details on skeleton diagrams. In \S2.2, we recall the essential details
of the vacuum moduli spaces of $\mathcal{N}=2$ theories. Next, in \S2.3,
we remind ourselves of the role of BPS quivers in these theories. In \S2.4, we consider
many of the important graphs which can be drawn on the Gaiotto curve $\mathcal{C}$.
In \S2.5, we introduced a technical definition of dessins d'enfants and the associated
Belyi maps. Finally, in \S2.6 we introduce the modular group and some important subgroups.

As a prelude to our discussion in \S3 of the connections between all these
objects, we provide in Figure 2 a roadmap of these connections, to aid orientation in the ensuing discussion.
 The first half
of \S3 will focus on the horizontal chain of correspondences shown
in Figure 2. Though most of these links
have been detailed in the literature previously,
it should be useful to present a codified story in one place,
with the focus on going from a skeleton diagram to the BPS spectrum
and Seiberg-Witten curve of that theory. Doing so will allow us to provide an extended discussion of the role
of dessins in these theories; this we shall do in the second half of \S3.


\begin{figure}

\begin{minipage}[t]{1\columnwidth}%
\noindent \begin{center}
\includegraphics{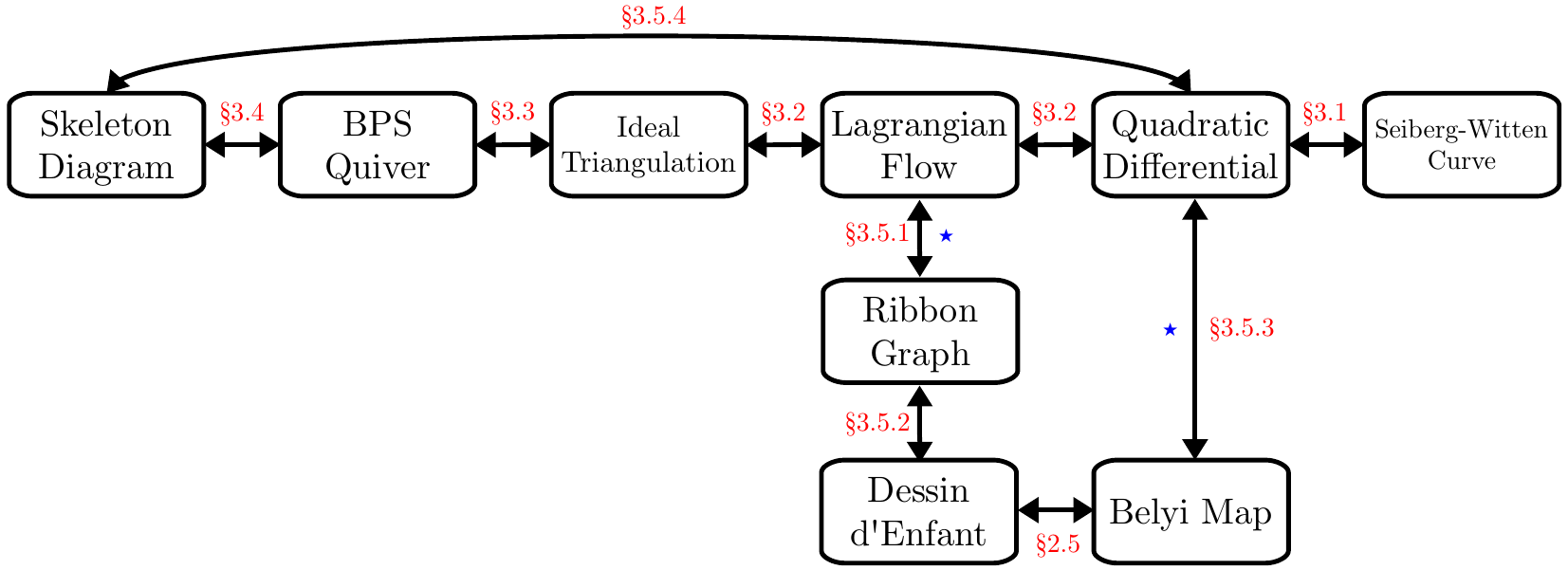}
\par\end{center}

\sf{\caption{A roadmap to the major connections between the objects of importance in the study of $SU\left(2\right)$ Gaiotto theories. Each of the major constructions of importance to
$SU\left(2\right)$ Gaiotto theories is shown. Arrows represent connections
between these entities. The section
in which each connection is discussed is indicated in red. Correspondences marked
with a blue star hold only when the quadratic differential on $\mathcal{C}$ is a Strebel differential; this shall 
be explained in \S3.}}

\end{minipage}
\end{figure}

\subsection{Skeleton Diagrams}

In \cite{Gaiotto}, Gaiotto found a new and interesting class of $\mathcal{N}=2$ supersymmetric
gauge theories in $\left(3+1\right)$ dimensions, obtainable from the wrapping of M5 branes over Riemann
surfaces. Following \cite{Hanany}, let us focus on the case where the gauge group is
a product of only $SU\left(2\right)$ factors. In this case, we can unambiguously represent the relevant
gauge theories as so-called \textbf{skeleton diagrams}, consisting of lines and trivalent nodes,
where a line represents an $SU\left(2\right)$ gauge group and a trivalent node represents
a matter field in the tri-fundamental representation $SU\left(2\right)^{3}$. Hence, these diagrams
can be seen as generalisations of the more familiar \textbf{quiver diagrams}, which have arisen
both in representation theoretic \cite{McKay} and gauge theoretic contexts
 \cite{Non-Abelian}. Indeed, whereas
fields charged under \emph{two} $SU$-factors, being the fundamental under one and the
anti-fundamental of another, readily afford description in terms of arrows in a quiver, fields charged
under more than two factors, as in our present case, require encoding beyond a quiver diagram.

Our skeleton diagrams are straightforward: they give rise to an infinite class of $\mathcal{N}=2$
gauge theories, having each line representing an $SU\left(2\right)$ gauge group with its
length inversely proportional to its gauge coupling $g_{YM}^2$ and each trivalent node
representing a half-hypermultiplet $Q_{\alpha\beta\gamma}$ transforming under the
tri-fundamental $\left(\square,\square,\square\right)$ representation of
$SU\left(2\right)\times SU\left(2\right)\times SU\left(2\right)$ with $\alpha,\beta,\gamma = 1,2$
indexing each of the $SU\left(2\right)$ factors. Any line of infinite length gives zero coupling
and the associated $SU\left(2\right)$ factor becomes a global flavour symmetry.
Because $\mathcal{N}=2$ supersymmetry is large enough to have its matter content determine
the interactions completely, each skeleton diagram thus defines a unique $\left(3+1\right)$-dimensional
$\mathcal{N}=2$ gauge theory. Any legitimate
skeleton diagram can be constructed simply through gluing together trivalent
vertices.

The string-theoretic realisation of $\mathcal{N}=2$ Gaiotto theories is in terms of a stack
 of M5 branes (for our $SU\left(2\right)$ Gaiotto theories, we have two M5 branes) wrapping
  a Riemann surface
$\mathcal{C}$ of genus $g$ with $n$ punctures. More precisely, consider M-theory in eleven dimensions
with coordinates $x^{0,\ldots,10}$ with $x^{7,8,9}=0$ fixed and $x^{0,1,2,3}$ the coordinates
of our four-dimensional world $\mathbb{R}^{4}_{x^{0,1,2,3}}$. Of the remaining four directions,
$Q_{x^{4,5,6,10}}\simeq \mathbb{R}^3 \times \mathbb{R}$, define a complex structure
$v=x^{4}+ix^{5}$ and $t=\mathrm{exp}\left(-\left(x^{6}+ix^{10}\right)/R_{IIA}\right)$ (so that
the $x^{10}$ direction indeed becomes periodic when compactifying to type IIA string theory
on a circle of radius $R_{IIA}$), and define a Riemann surface 
$\mathcal{C}=\left\{ F\left(v,t\right)=0\right\} \subset Q$ over which the M5 brane can wrap.
In the type IIA perspective, this corresponds to $n+1$ NS5-branes occupying $x^{0,1,\ldots,5}$ and
placed in parallel at fixed values of $x^{6}$; moreover, between adjacent pairs of
NS5-branes are stretched stacks of D4-branes. The variable names are chosen judiciously:~the
skeleton diagram of the theory to which $\mathcal{C}$ corresponds is one whose topology,
graphically, consists of $g$ independent closed circuits and $n$ external (semi-infinite) legs.

We can easily check \cite{Hanany} that given a skeleton diagram specified by the pair 
$\left\langle g,n\right\rangle $, the number of internal (finite) lines, hence the number of
$SU\left(2\right)$ gauge group factors, is $3g-3+n$, while the number of nodes, hence the number
of matter fields, is the Euler characteristic $2g-2+n$.
The Coulomb branch of the moduli
space of the Gaiotto theory in question (to be discussed in more depth below)
is specified by the topology $\left\langle g,n\right\rangle $
of the skeleton diagram \cite{Hanany,HeJohn}. Finally, it is worth noting 
that the only features of the skeleton diagrams
of these theories relevant to the physics they encode are the parameters
$g$ and $n$: this is sometimes referred to as
\textbf{Gaiotto duality} \cite{CecottiVafa}.

\subsection{Moduli Spaces}

For $\mathcal{N}=2$ theories, the space of vacuum expectation values
of the theory is known as the vacuum \textbf{moduli space} of the
theory. One can think of this as the space of minima of an effective
potential, governed by the zero loci of a set of algebraic equations.
Thus the moduli space is an \emph{affine algebraic variety }on a
complex space $\mathbb{C}^{k}$, whose coordinates are the vacuum
expectation values. As a variety, the moduli space of an
$\mathcal{N}=2$ theory typically decomposes into two \emph{branches},
known as the \emph{Higgs branch }$\mathcal{B}$ and \emph{Coulomb
branch }$\mathcal{U}$. The former is parameterised
by the massless gauge singlets of the hypermultiplets, occurring where
the gauge group is completely broken and the vector multiplet becomes
massive via the Higgs mechanism. The latter is parameterised by the
complex scalars in the vector multiplet, occurring when the gauge
group is broken to some Abelian subgroup and the hypermultiplets generically
become massive.

Though our focus in this paper will mostly be on the Coulomb branch $\mathcal{U}$,
it is worth noting at this point that for $g>0$, where there is more than a single $SU\left(N\right)$
factor, the gauge group may not be completely broken on the Higgs branch of each factor
and thus to avoid confusion, the authors of \cite{Hanany} dub this quasi-Higgs branch the
\emph{Kibble branch} $\mathcal{K}$. A beautiful result of \cite{Hanany} is that the Kibble branch
of the moduli space is an algebraic variety such that

\begin{equation}
\mathrm{dim}_{\mathbb{H}}\left(\mathcal{K}\right)=n+1 ,
\end{equation}

\noindent where $\mathrm{dim}_{\mathbb{H}}$ means the \emph{quaternionic} dimension, i.e.~four
 times the real dimension or twice the complex dimension. It is interesting to see that this result
is independent of $g$. A quick argument would proceed as follows: each trivalent node consists
of 4 quaternionic degrees of freedom and there are $\chi = 2g - 2 +n$ thereof; generically on
$\mathcal{K}$, the $SU\left(2\right)^{3g-3+n}$ gauge group breaks to $U\left(1\right)^{g}$, hence
$3\left(3g-3+n\right)-g$ broken generators. Thus, there are effectively
 $4\chi-\left(3\left(3g-3+n\right)-g\right)=n+1$ quaternonic degrees of freedom.

\subsection{BPS Quivers}

$\mathcal{N} = 2$ Gaiotto theories admit \textbf{BPS quivers }\cite{CecottiVafa}.
Let us briefly recall the details of these diagrams, following the discussion
in \cite{FirstVafa}. We begin with a  $\left(3+1\right)$ dimensional $\mathcal{N}=2$
theory with Coulomb moduli space $\mathcal{U}$.
At a generic point $u\in\mathcal{U}$, the theory has $U\left(1\right)^{r}$
gauge symmetry, and a low-energy solution defined by:
\begin{itemize}
\item A lattice $\Gamma$ of electric, magnetic and flavour charges.
\item An antisymmetric inner product on the charge lattice $\circ:\Gamma\times\Gamma\rightarrow\mathbb{C}$.
\item A linear function $\mathcal{Z}_{u}:\Gamma\rightarrow\mathbb{C}$,
the \emph{central charge function} of the theory.
\end{itemize}
The central charge function $\mathcal{Z}_{u}$ naturally appears in
the $\mathcal{N}=2$ algebra, and provides a lower bound on the masses
of charged particles. The mass of a particle with $\gamma\in\Gamma$
satisfies $M\geq\left|\mathcal{Z}_{u}\left(\gamma\right)\right|$.
The lightest charged particles are those that saturate this bound
- these are termed \textbf{BPS states}.

The BPS quiver allows computation of the full BPS spectrum of the
theory at some fixed point $u$ in the Coulomb branch, supposing that
the occupancy of the BPS states at that point is known \cite{MainVafa}.
This dramatically simplifies the problem of finding BPS states, since
in place of some tedious weak coupling physics or prohibitively difficult strong coupling
dynamics, the BPS spectrum is governed by a quantum mechanics problem encoded
in the BPS quiver \cite{MainVafa}.
 
To construct a BPS quiver, first choose a half-plane $\mathcal{H}$
in the complex $\mathcal{Z}_{u}$ plane. All states with central charge
in $\mathcal{H}$ will be considered particles, while the states in
the opposite half-plane will be considered anti-particles. Suppose
there exists a set of hypermultiplet states $\left\{ \gamma_{i}\right\} $
in the chosen half-plane that forms a positive integral basis for
all particles. Given this basis $\left\{ \gamma_{i}\right\} $, we
can construct a BPS quiver as follows: For every charge $\gamma_{i}$,
draw a node associated to it. For every pair of charges $\gamma_{i}$,
$\gamma_{j}$ with $\gamma_{i}\circ\gamma_{j}>0$, draw $\gamma_{i}\circ\gamma_{j}$
arrows from $\gamma_{i}$ to $\gamma_{j}$. The importance of the
BPS quiver is that it can be used to check whether a particular site
of the charge lattice $\gamma=\sum_{i}n_{i}\gamma_{i}\in\Gamma$ is
occupied by a BPS state, and if so, to determine the spin and degeneracy
of the associated particles. If we do this for every $\gamma$, we
will have computed the full BPS spectrum of the theory at $u\in\mathcal{U}$
(for the details here, the reader is 
referred to \cite{MainVafa,FirstVafa}).

Importantly, $\mathcal{Z}_{u}$ varies from point to point in $\mathcal{U}$,
hence its subscript $u$. A consequence of this is that the above
procedure can yield different BPS quivers at different points $u\in\mathcal{U}$,
thereby partitioning $\mathcal{U}$ into domains, each corresponding
to a different BPS quiver for the theory. For an $\mathcal{N}=2$
Gaiotto theory, there is always a finite number of such BPS quivers
\cite{CecottiVafa}. Together, they are said to form a \emph{finite
mutation class}, and there exists an algorithm, known as the \emph{mutation
method}, to enumerate all BPS quivers in a mutation class, once one
has been specified.

Roughly, the principle on which the mutation method is based is as
follows.
Up to this point, we have arbitrarily chosen
a half-plane $\mathcal{H}$ in the complex $\mathcal{Z}_{u}$ plane
when constructing the BPS quiver. This choice of half-plane yields
a unique basis of BPS states $\left\{ \gamma_{i}\right\} $ and corresponding
BPS quiver. Now consider rotating the half-plane clockwise by an angle
$\theta$, so that $\mathcal{H}\rightarrow\mathcal{H}_{\theta}=e^{-i\theta}\mathcal{H}$.
As we tune $\theta$ from zero, nothing happens while all the $\left\{ \gamma_{i}\right\} $
remain in the half-plane $\mathcal{H}_{\theta}$. However, for some
value of $\theta$, the left-most state $\gamma_{1}$ will exit $\mathcal{H}_{\theta}$
on the left, while simultaneously the antiparticle state $-\gamma_{1}$
will enter on the right. Thus we will have a new basis of elementary
BPS states $\left\{ \tilde{\gamma}_{i}\right\} $ and a corresponding
new BPS quiver. There is a simple algorithm for constructing this
new basis of BPS states and corresponding BPS quiver from the old
\cite{MainVafa,CecottiVafa,FirstVafa}. One full rotation of the half
plane will enumerate the full mutation class of BPS quivers for the
theory. It is important to remember that all the BPS quivers in a mutation
 class for a specific Gaiotto theory
encode exactly the same physics \cite{MainVafa,CecottiVafa,FirstVafa}.


\subsection{Quadratic Differentials and Graphs on Gaiotto Curves}

A \textbf{quadratic differential} on a Riemann surface $S$ is a map

\begin{equation}
\phi : TS \rightarrow \mathbb{C}
\end{equation}

\noindent satisfying $\phi\left(\lambda v\right) = \lambda^2 \phi \left(v\right)$ for all $v \in TS$ and all $\lambda \in \mathbb{C}$. If $z: U \rightarrow \mathbb{C}$ is a chart defined on some open set $U \subset S$, then $\phi$ is equal
on $U$ to

\begin{equation}
\phi_U \left(z\right) \mathrm{d}z^2
\end{equation}

\noindent for some function $\phi_U$ defined on $z\left(U\right)$. Now suppose that two charts $z:U\rightarrow \mathbb{C}$ and $w:V\rightarrow \mathbb{C}$ on $S$ overlap, with transition function $h:=w \circ z^{-1}$; then if
$\phi$ is represented both as $\phi_U \left(z\right) \mathrm{d}z^2$ and $\phi_V \left(w\right) \mathrm{d}w^2$ on $U \cap V$, we have \cite{Strebel}

\begin{equation}
\phi_V \left(h\left(z\right)\right) \left(h'\left(z\right)\right)^2 = \phi_U\left(z\right) .
\end{equation}

With these basic facts in mind, let us return to the  $SU\left(2\right)$ Gaiotto theories of interest. Here, the Seiberg-Witten curves for these theories
have the form $y^{2}=\phi\left(x\right)$, where $q=\phi\left(x\right)\mathrm{d}x^{2}$
is a {quadratic differential} on $\mathcal{C}$ with only second
order poles \cite{Gaiotto,MainVafa}. (A subtlety:~double poles in the quadratic differential only emerge for a mass-deformed theory. For vanishing mass deformation, the poles of the differential are single poles.) The functional form of $q$ is specified
by the topology $\left\langle g,n\right\rangle $ of the skeleton
(we shall elaborate the exact details of this specification in the following section),
and varying individual parameters in $q$ amounts to changing the
point in the moduli space under consideration. Since the Gaiotto theory in question is specified by $\left\langle g,n\right\rangle $, the Coulomb branch for each theory is determined by these two parameters; for this reason, we shall sometimes write $\mathcal{U}_{g,n}$ for the Coulomb branch of the Gaiotto theory with $\left\langle g,n\right\rangle $. In addition, $q$ depends on $n$ further positive real parameters associated to the $n$ punctures of $\mathcal{C}$ (which correspond to masses, couplings and moduli not fixed by moving in the Coulomb branch); the details of this dependence shall again be spelt out in the following section. To completely fix $q$ requires us to fix a point in $\mathcal{U}_{g,n} \times \mathbb{R}^n_+$.

As we shall demonstrate
in \S3, the BPS quiver for a specific $\mathcal{N}=2$ theory at a
point $u\in\mathcal{U}$ can be constructed from a specific quadratic
differential $q$ on $\mathcal{C}$ \cite{MainVafa}, by using the quadratic differential to construct
a graph on $\mathcal{C}$ known as its \textbf{ideal triangulation}, which has marked points as
nodes and zeroes as faces. There
is also a translation between BPS quivers and skeleton diagrams, which
we shall again elaborate in \S3 of this paper.

\subsection{Dessins d'Enfants and Belyi Maps}

There is a connection between the quadratic differentials on $\mathcal{C}$
and Grothendieck's \textbf{dessins d'enfants} \cite{DessinBook,LMS1,LMS2,LMS3}. Such a dessin is an
ordered pair $\left\langle X,\mathcal{D}\right\rangle $ where $X$
is an oriented compact topological surface (here the Gaiotto curve
$\mathcal{C}$) and $\mathcal{D}\subset X$ is a finite graph\textbf{
}satisfying the following conditions \cite{DessinBook}:
\begin{enumerate}
\item $\mathcal{D}$ is connected.
\item $\mathcal{D}$ is \emph{bipartite}, i.e.~consists of only black and
white nodes, such that vertices connected by an edge have different
colours.
\item $X\setminus\mathcal{D}$ is the union of finitely many topological
discs, which we call the \emph{faces} of $\mathcal{D}$.
\end{enumerate}

As we shall show in \S3, at certain points in the Coulomb branch (and fixing in addition parameters associated to masses and couplings of the global flavour symmetries) we can use the quadratic differential to construct a graph
on $\mathcal{C}$ known as a \textbf{ribbon graph} \cite{Mulase}, with marked points as faces and zeroes
as nodes. At these points in
the moduli space, the quadratic differential satisfies the conditions to be a so-called \textbf{Strebel differential}.
As we shall see, we can interpret the ribbon graphs (with edge lengths set to be equal -- corresponding to a \emph{specific} fixing of the parameters associated to the flavour symmetries) as dessins by inserting a coloured
node into every edge and colouring every vertex white; doing so leads to a number of interesting mathematical ramifications which cement
dessins as important objects of study in the context of these $\mathcal{N}=2$ theories (all of which shall be
discussed in depth in the following section).
In addition, we note here that if all the nodes of one of the two possible colours have valency two,
then the dessin in question is referred to as \emph{clean} \cite{DessinBook}.

Now recall that there is a one-to-one correspondence
between dessins d'enfants and \textbf{Belyi maps} \cite{DessinBook,JS2}.
A Belyi map is a holomorphic map to $\mathbb{P}^{1}$ ramified at
only $\left\{ 0,1,\infty\right\} $, i.e. for which the only points
$\tilde{x}$ where $\frac{\mathrm{d}}{\mathrm{d}x}\beta\left(x\right)|_{\tilde{x}}=0$
are such that $\beta\left(\tilde{x}\right)\in\left\{ 0,1,\infty\right\} $.
We can associate a Belyi map $\beta\left(x\right)$ to a dessin via
its \emph{ramification indices}: the order of vanishing of the Taylor
series for $\beta\left(x\right)$ at $\tilde{x}$ is the ramification
index $r_{\beta\left(\tilde{x}\right)\in\left\{ 0,1,\infty\right\} }\left(i\right)$
at that $i$th ramification point \cite{HeJohn,YMR}. To draw the dessin
from the map, we mark one white node for the $i$th pre-image of 0,
with $r_{0}\left(i\right)$ edges emanating therefrom; similarly,
we mark one black node for the $j$th pre-image of 1, with $r_{1}\left(j\right)$
edges. We connect the nodes with the edges, joining only black with
white, such that each face is a polygon with $2r_{\infty}\left(k\right)$
sides (see e.g.~\cite{YMR, LMS1}).

\subsection{The Modular Group and Congruence Subgroups}

Finally, we should very briefly recall some essential details regarding the modular group
$\Gamma\equiv\mathrm{\Gamma\left(1\right)=PSL}\left(2,\mathbb{Z}\right)=\mathrm{SL}\left(2,\mathbb{Z}\right)/\left\{ \pm I\right\} $.
This is the group of linear fractional transformations $\mathbb{Z} \ni z\rightarrow\frac{az+b}{cz+d}$, 
with $a,b,c,d\in\mathbb{Z}$ and $ad-bc=1$. It is generated by the
transformations $T$ and $S$ defined by:
\begin{equation}
T(z)=z+1\quad,\quad S(z)=-1/z \ .
\end{equation}
The presentation of $\Gamma$ is $\left\langle S,T|S^{2}=\left(ST\right)^{3}=I\right\rangle$.

The most important subgroups of $\Gamma$ are the
so-called \emph{congruence} subgroups, defined by having the the entries in the generating matrices $S$ and $T$ obeying some modular arithmetic. Some conjugacy classes of congruence subgroups of particular
note are the following:

\begin{itemize}
\item Principal congruence subgroups:
\[
\Gamma\left(m\right):=\left\{ A\in\mathrm{SL}(2;\mathbb{Z})\;;\; A\equiv\pm I\;\mathrm{mod}\; m\right\} /\left\{ \pm I\right\} ;
\]

\item Congruence subgroups of level $m$: subgroups of $\Gamma$ containing
$\Gamma\left(m\right)$ but not any $\Gamma\left(n\right)$ for $n<m$;
\item Unipotent matrices:
\[
\Gamma_{1}\left(m\right):=\left\{ A\in\mathrm{SL}(2;\mathbb{Z})\;;\; A\equiv\pm\begin{pmatrix}1 & b\\
0 & 1
\end{pmatrix}\;\mathrm{mod}\; m\right\} /\left\{ \pm I\right\} ;
\]

\item Upper triangular matrices:
\[
\Gamma_{0}\left(m\right):=\left\{ \begin{pmatrix}a & b\\
c & d
\end{pmatrix}\in\Gamma\;;\; c\equiv0\;\mathrm{mod}\; m\right\} /\left\{ \pm I\right\} .
\]

\end{itemize}

In \cite{HeJohn}, our attention is drawn to the conjugacy classes of a particular family of subgroups
of $\Gamma$:~the so-called\emph{ genus zero, torsion-free} congruence
subgroups. 
By \emph{torsion-free} we mean that the subgroup
contains no element of finite order other than the identity. 
To explain
\emph{genus zero}, first recall that the modular group acts on the
upper half-plane $\mathcal{H}:=\left\{ \tau\in\mathbb{C}\;,\;\mathrm{Im}\left(\tau\right)>0\right\} $
by linear fractional transformations $z\rightarrow\frac{az+b}{cz+d}$.
$\mathcal{H}$ gives rise to a compactification $\mathcal{H}^{*}$ when adjoining
\emph{cusps}, which are points on $\mathbb{R}\cup\infty$ fixed
under some parabolic element (i.e.~an element $A\in\Gamma$ not equal to the identity
 and for which $\mathrm{Tr}\left(A\right)=2$). The quotient $\mathcal{H}^{*}/\Gamma$
is a compact Riemann surface of genus 0, i.e.~a sphere. It turns out
that with the addition of appropriate cusp points, the extended upper
half plane $\mathcal{H}^{*}$ factored by various congruence subgroups
will also be compact Riemann surfaces, possibly of higher genus. Such
a Riemann surface, as a complex algebraic variety, 
is called a {\bf modular curve}. 
The genus of a subgroup of the modular group is the genus of the modular
curve produced in this way. The conjugacy classes of the genus zero
 torsion-free congruence subgroups of the modular group are very rare:~there are only 33 of them,
 with index $I\in\left\{ 6,12,24,36,48,60\right\} $. \cite{MS}


\section{A Web of Correspondences}

In this section, we elaborate the chain of connections between skeleton
diagrams, BPS quivers, ideal triangulations, ribbon graphs, quadratic
differentials and Seiberg-Witten curves for $SU\left(2\right)$ Gaiotto
theories. In \S3.1, we present the connection between the quadratic
differential $q=\phi\left(x\right)\mathrm{d}x^{2}$ and Seiberg-Witten
curve $y^{2}=\phi\left(x\right)$ for these theories \cite{Gaiotto}. In
\S3.2, we show how the quadratic differential on $\mathcal{C}$ encodes
the \emph{special Lagrangian flow} on this Riemann surface. We
then describe how the ideal triangulation on $\mathcal{C}$ is
obtained. 
In \S3.3, we recall how to construct a BPS
quiver from a certain ideal triangulation \cite{MainVafa}.
Next, in \S3.4 we
show how to translate between skeleton diagrams and BPS quivers. 
In \S3.5, we show how the special Lagrangian flow on $\mathcal{C}$
can be used, at certain special points in the Coulomb branch where the quadratic differential becomes Strebel,
 to construct a ribbon graph on $\mathcal{C}$, and from this describe in detail how dessins d'enfants arise in the 
 context of these theories, and the important roles they play in their study.


\subsection{Quadratic Differentials and Seiberg-Witten Curves}

The physics of an $SU\left(2\right)$ Gaiotto theory is determined
by a Riemann surface $\mathcal{C}$ of genus $g$ with $n$ punctures,
one at each marked point $p_{i}\in\mathcal{C}$ \cite{Gaiotto}. We select
a particular meromorphic quadratic differential $q=\phi\left(x\right)\mathrm{d}x^{2}$
on $\mathcal{C}$.
Fixing the behaviour of $q$ at the points $p_{i}$ by admitting a pole of finite order
amounts to imposing that near $p_{i}$:

\begin{equation}
q\left(x\right)\sim\frac{1}{x^{k_{i}+2}}\mathrm{d}x^{2}
\end{equation}

The integer $k_{i}\geq0$ associated to each puncture is invariant
under changes of coordinates. For the $SU\left(2\right)$ Gaiotto theories in question, 
we always choose  $k_{i}=0$ \cite{Gaiotto,MainVafa}. The Seiberg-Witten curve $\Sigma$ of
the theory is given by a double cover of $\mathcal{C}$, and we obtain
the \textbf{Seiberg-Witten differential} $\lambda$ as follows \cite{MainVafa}:

\begin{equation}
\Sigma=\left\{ \left(x,y\right)|y^{2}=\phi\left(x\right)\right\} ,\qquad\lambda=y\mathrm{d}x=\sqrt{q}
\end{equation}

\noindent Note that by varying the quadratic differential we obtain a family
of Seiberg-Witten curves, and in this way the Coulomb branch $\mathcal{U}_{g,n}$
of the theory is naturally identified with the space of quadratic
differentials obeying the boundary conditions from equation (3.1), \emph{up to} a dependence on $n$ further positive real parameters associated to the $n$ punctures of $\mathcal{C}$
(these numbers will turn out to be related to masses, couplings and moduli of the global flavour symmetries
of the theory, not fixed by moving in the Coulomb branch). Thus, to completely fix $q$, we must fix a point in $\mathcal{U}_{g,n} \times \mathbb{R}^n_+$ \cite{MainVafa, Mulase}.

\subsection{Trajectories on Riemann Surfaces and Ideal Triangulations}

Consider a Riemann surface $\mathcal{C}$ with a meromorphic quadratic
differential $q$. Locally, we can write $q=\phi\left(x\right)\mathrm{d}x^{2}$
for the appropriate local coordinate $x$ on $\mathcal{C}$. Using
this quadratic differential, we can classify parametric curves $\gamma\left(t\right)$
on $\mathcal{C}$ according to the sign of $q$. \emph{Horizontal
trajectories} are defined as those for which $\phi\left(\gamma\left(t\right)\right)\dot{\gamma}\left(t\right)^{2}>0$,
while \emph{vertical trajectories} are defined as those for which
$\phi\left(\gamma\left(t\right)\right)\dot{\gamma}\left(t\right)^{2}<0$
\cite{Mulase,Strebel}. There are three cases of importance in the
study of these trajectories:~a generic point on $\mathcal{C}$, a
zero of $q$ on $\mathcal{C}$, and a pole of $q$ on $\mathcal{C}$.

Let us first consider a generic point on $\mathcal{C}$. To begin,
suppose that $q=\mathrm{d}x^{2}$. Then the horizontal trajectories
are given by the horizontal lines $\alpha\left(t\right)=t+ci$ and
the vertical trajectories are given by the vertical lines $\beta\left(t\right)=it+c$
\cite{Strebel}. Now, if a quadratic differential $q=\phi\left(x\right)\mathrm{d}x^{2}$
is holomorphic and non-zero at $x=x_{0}$, then on a neighbourhood
of $x_{0}$ we can introduce the canonical coordinate $w\left(x\right)=\int_{x_{0}}^{x}\sqrt{\phi\left(x\right)}\mathrm{d}x$.
It follows from the transformation rule (2.4) that in terms of the
canonical coordinate the quadratic differential is given by $q=\mathrm{d}w^{2}$,
so at a generic point on $\mathcal{C}$ the horizontal and vertical
trajectories are horizontal and vertical lines, as shown in Figure
3(a) \cite{Mulase}.

The situation is different where $q$ either vanishes or has a pole.
Let us consider what the horizontal and vertical trajectories look
like in the vicinity of such points, which without loss of generality
we shall take to be at zero. First, suppose that $q$ vanishes here,
so that $q=x^{m}\mathrm{d}x^{2}$. Then, with $t\in\mathbb{R}^{+}$,
the horizontal trajectories are given by $\left(m+2\right)$ half-rays
that have $x=0$ on the boundary \cite{Mulase}:

\begin{equation}
\alpha_{k}\left(t\right)=t\cdot\exp\left(\frac{2\pi ik}{m+2}\right),\qquad k=0,1,\ldots,m+1 .
\end{equation}

\noindent The vertical trajectories are given by another set of $\left(m+2\right)$
half-rays that have $x=0$ on the boundary \cite{Mulase}:

\begin{equation}
\beta_{k}\left(t\right)=t\cdot\exp\left(\frac{\pi i+2\pi ik}{m+2}\right),\qquad k=0,1,\ldots,m+1 .
\end{equation}

\noindent Thus we see that in the neighbourhood of zero both types of trajectory
look like rays emanating from zero at some discrete angles, as shown
in Figure 3(b). For a so-called \emph{simple zero}, we have $m=1$,
and these trajectories make angles of $2\pi/3$ with each other \cite{MainVafa}.

Now consider the case where $q$ has a second order pole. We take
$q=-x^{-2}\mathrm{d}x^{2}$. Then, horizontal trajectories are concentric
circles centered at zero \cite{Mulase}:

\begin{equation}
\alpha\left(t\right)=re^{it},\qquad t\in\mathbb{R},\quad r>0 .
\end{equation}

\noindent Vertical trajectories are given by half-rays emanating from zero \cite{Mulase}:

\begin{equation}
\beta\left(t\right)=te^{i\theta},\qquad t>0,\quad0\leq\theta<2\pi .
\end{equation}

\noindent These trajectories in the vicinity of a second order pole are hence
as shown in Figure 3(c).

\begin{figure}

\begin{center}
\begin{minipage}[t]{0.25\textwidth}%
\begin{center}
\includegraphics{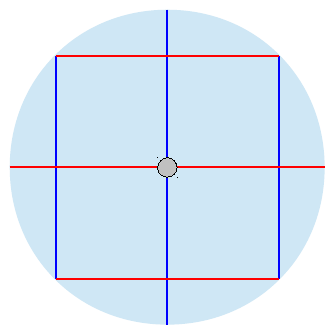}
\par
(a)\end{center}
%
\end{minipage}\qquad{}%
\begin{minipage}[t]{0.25\textwidth}%
\begin{center}
\includegraphics{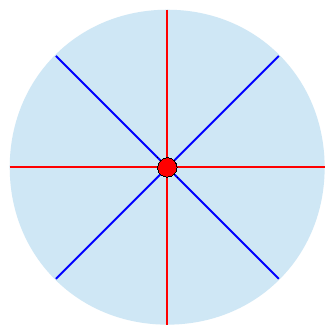}
\par
(b)\end{center}
%
\end{minipage}\qquad{}%
\begin{minipage}[t]{0.25\textwidth}%
\begin{center}
\includegraphics{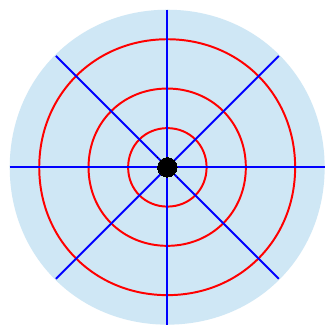}
\par
(c) \end{center}
%
\end{minipage}
\par\end{center}

\sf{\caption{Horizontal (red) and vertical (blue)
trajectories in the neighbourhood of (a) a generic point (marked in grey), (b)
a zero of $q$ (marked in red), and (c) a second order pole of $q$ (marked in black)
on $\mathcal{C}$.}}
\end{figure}

Note that we have only discussed second order poles. In \cite{MainVafa},
the analysis of the punctures is split into two cases depending on
the order $k_{i}+2$ of the pole in $q$. The \emph{regular punctures}
in $\mathcal{C}$ are those for which $k_{i}=0$. By Gaiotto's prescription,
these are associated with flavour symmetries \cite{Gaiotto}. The \emph{irregular
punctures} in $\mathcal{C}$ are those for which $k_{i}>0$. For our
purposes, we need consider only regular punctures:~irregular punctures
are associated with boundaries of $\mathcal{C}$, and none of the
Gaiotto curves for the $SU\left(2\right)$ theories of interest have
boundaries  \cite{CecottiVafa,Gaiotto}.

 The trajectories described define the \emph{special Lagrangian
flow lines} on $\mathcal{C}$ \cite{MainVafa}. Suppose that we have a
surface $\mathcal{C}$ with $n$ marked points where $q$ develops
a second order pole. As we just saw, the horizontal trajectories are
concentric rings around the marked points. This defines \emph{domains}
for each point. These domains are separated by the radial lines going
between different zeros of $q$. Only at very special points in the Coulomb branch $\mathcal{U}_{g,n}$ will
these trajectories be such that they define
a graph drawn on $\mathcal{C}$, where the marked points can be identified
with the faces. Such a graph is known as a \textbf{ribbon graph} \cite{Mulase};
an extended discussion of such graphs and the circumstances in which they can be drawn is postponed to \S3.5.%
{} It turns out that there are six topologically distinct possible ribbon graphs 
for a Gaiotto theory with one $SU\left(2\right)$
factor and $N_{f}=4$ flavours; one example is drawn in Figure 4(a)
 (in this case, we know from \cite{Gaiotto} that the quadratic differential
has the form $\phi=P_{4}\left(x\right)/\Delta_{4}^{2}\left(x\right)$,
where $P$ and $\Delta$ are polynomials in $x$ and subscripts indicate
polynomial degrees); the rest are drawn (as dessins) in Figure 6.

A convenient way to encode the topological structure of the special Lagrangian flow
 is in an \textbf{ideal triangulation} of \emph{$\mathcal{C}$.}
To construct such a triangulation, we consider one generic flow line
which has its endpoints at two marked points on $\mathcal{C}$, and
connect those two marked points by that trajectory. We repeat this
for all pairs of marked points which are connected by such generic
trajectories \cite{MainVafa}. The structure of the flow on $\mathcal{C}$
is such that each face of the resulting graph will have three edges
and contain exactly one zero of $q$ (in general we assume that these
zeroes are simple, and thus have three radially outgoing trajectories,
as discussed above) \cite{MainVafa}. To illustrate, an ideal triangulation
for a Gaiotto theory with one $SU\left(2\right)$ factor and $N_{f}=4$
flavours is drawn in Figure 4(b).

\begin{figure}

\begin{center}
\begin{minipage}[t]{0.47\textwidth}%
\begin{center}
\includegraphics{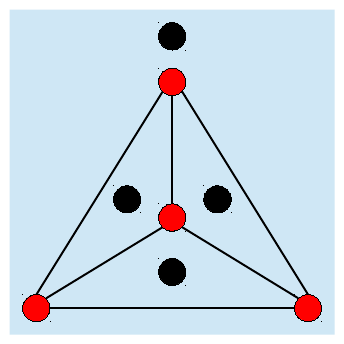}
\par
(a)\end{center}

%
\end{minipage}\qquad{}%
\begin{minipage}[t]{0.47\textwidth}%
\begin{center}
\includegraphics{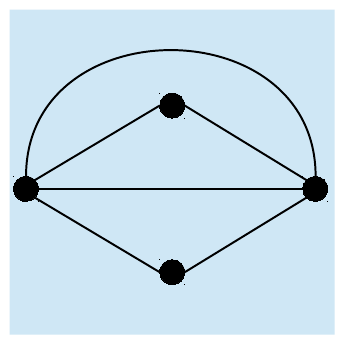}
\par
(b)\end{center}

%
\end{minipage}
\par\end{center}

\sf{\caption{Ribbon graph (a) and corresponding ideal triangulation (b) for the 
$SU\left(2\right)$, $N_{f}=4$ theory. Black nodes denote punctures of $\mathcal{C}$;
red nodes denote zeroes of $q$. The ideal triangulation has four faces, for the 
four zeroes of $q$.}}

\end{figure}

 
 In a sense, the ideal triangulation can be considered ``dual'' to the ribbon graph, as while an ideal triangulation has marked points as vertices and zeroes of the quadratic differential as faces, the opposite is true for a ribbon graph. However, caution is needed here, as while it is generically true at \emph{any} point in the Coulomb branch $\mathcal{U}$ that one can construct an ideal triangulation on $\mathcal{C}$, one can \emph{only} construct ribbon graphs, via the prescription involving horizontal and vertical trajectories above, at the \emph{specific} points in $\mathcal{U}$ (which will turn out to be the points where the quadratic differential on $\mathcal{C}$ satisfies the stricter definition of being a \emph{Strebel differential}, as discussed in section \ref{s:ribbon}). The picture that emerges is therefore as follows. At any point in $\mathcal{U}$, one can construct an ideal triangulation. To each such ideal triangulation it is possible to associate a BPS quiver (see section \ref{s:BPS}). There are finitely many BPS quivers, so $\mathcal{U}$ is partitioned into domains corresponding to each such BPS quiver. In addition, there are finitely many \emph{specific points} in $\mathcal{U}$ where one can construct a ribbon graph. 

\subsection{Constructing BPS Quivers}\label{s:BPS}

From the structure of an ideal triangulation on $\mathcal{C}$, there
is a simple algorithm to extract the corresponding BPS quiver. We
refer to an edge in the triangulation as a \emph{diagonal} $\delta$
if the edge does not lie on a boundary of $\mathcal{C}$. (For the
$SU\left(2\right)$ Gaiotto theories under consideration this makes
no difference, since all the Gaiotto curves in this case are without
boundary, as discussed.) The algorithm to construct a theory's BPS
quiver from its ideal triangulation on $\mathcal{C}$ is then as follows
\cite{MainVafa}:
\begin{itemize}
\item For each diagonal $\delta$ in the triangulation, draw one node of
the BPS quiver.
\item For each pair of diagonals $\delta_{1}$ and $\delta_{2}$ in the
triangulation, find all the triangles for which both specified diagonals
are edges. For each such triangle, draw one arrow connecting the nodes
defined by $\delta_{1}$ and $\delta_{2}$. Determine the direction
of the arrow by looking at the triangle shared by $\delta_{1}$ and
$\delta_{2}$. If $\delta_{1}$ immediately precedes $\delta_{2}$
going anti-clockwise around the triangle, the arrow points from $\delta_{1}$
to $\delta_{2}$.
\end{itemize}

For the derivation of this algorithm, the reader is referred to the original source
\cite{MainVafa}. It is straightforward to confirm that the BPS quiver
for the Gaiotto theory with one $SU\left(2\right)$ factor and $N_{f}=4$
flavours shown in Figure 1(b) can be constructed from the corresponding
triangulation shown in Figure 4(b).

\subsection{Skeleton Diagrams and BPS Quivers}

With these connections between BPS quivers, ideal triangulations,
flow diagrams, quadratic differentials on $\mathcal{C}$, and Seiberg-Witten
curves established
(following the details given in \cite{MainVafa,FirstVafa,CecottiVafa}),
it remains to see how the skeleton diagrams for
the corresponding $SU\left(2\right)$ Gaiotto theories enter the picture.
To do this, first recall that each puncture of $\mathcal{C}$ corresponds
to a global $SU\left(2\right)$ flavour symmetry. Any two Gaiotto
curves can be glued together by opening a hole at a puncture and gluing
the two together with a tube; this results in gauging
the $SU\left(2\right)$ groups corresponding to the two punctures
\cite{FirstVafa}. By following this procedure, any Gaiotto curve
can be constructed by gluing together three-punctured spheres; at
the level of the skeleton, this simply amounts to joining trivalent
vertices \cite{Hanany}.

In \cite{FirstVafa} it is shown that this gluing procedure for
the Gaiotto curves/skeleton diagrams can be translated into a gauging
rule for the BPS quivers. To gauge a symmetry, we add gauge degrees
of freedom and couple them to the matter already present in the theory.
At the level of the quiver, this amounts to adding two nodes of a
pure $SU\left(2\right)$ subquiver to add the gauge degrees of freedom,
then coupling the existing pairs of identical nodes corresponding
to the $SU\left(2\right)$ flavour symmetries to this subquiver.
To do this, we delete one of the two identical nodes in each case
and connect the other to the $SU\left(2\right)$ subquiver in an oriented
triangle; the deleted state will then be generated by a bound state
within the $SU\left(2\right)$ nodes \cite{FirstVafa}.

Suppose that we are now given the skeleton diagram for the four-punctured
sphere, as shown in Figure 1(a). This is the most basic legitimate skeleton diagram
(one internal edge corresponding to one $SU\left(2\right)$ factor, formed
by joining two trivalent vertices together); the corresponding BPS quiver was found in
\cite{MainVafa,CecottiVafa}, and is drawn in Figure 1(b). One can now construct
any other legitimate skeleton diagram by appending more trivalent vertices onto
any of the external legs; using the above procedure, one can construct a corresponding
BPS quiver in every case.
In this way, we obtain a precise translation between skeleton diagrams and
BPS quivers for the $SU\left(2\right)$ Gaiotto theory in question,
completing the backbone of correspondences in Figure 2. Once we have
obtained one such BPS quiver, the rest in its finite mutation class
can be computed using the mutation method \cite{MainVafa,CecottiVafa,FirstVafa}.
An example of this gauging procedure for the case of gauging two $SU\left(2\right)$,
$N_{f}=4$ Gaiotto theories is provided in Figure 5.

\begin{figure}

\begin{center}
\begin{minipage}[t]{0.47\textwidth}%
\begin{center}
\includegraphics{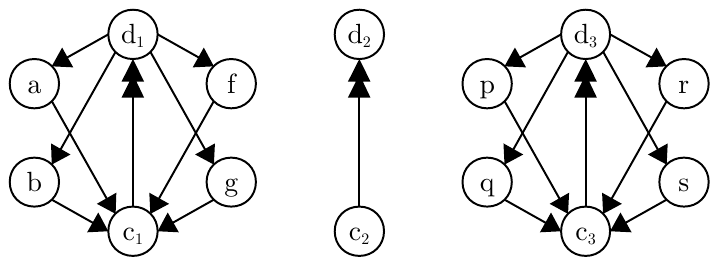}
\par
(a)\end{center}

\end{minipage}\qquad{}%
\begin{minipage}[t]{0.47\textwidth}%
\begin{center}
\includegraphics{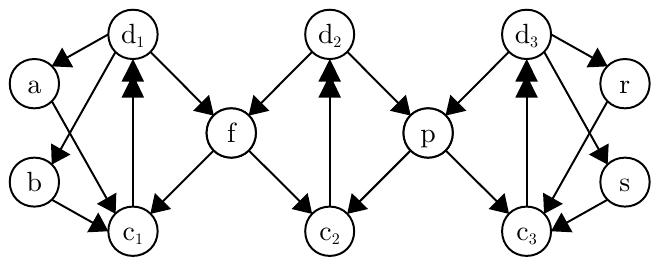}
\par
(b)\end{center}

\end{minipage}
\par\end{center}

\sf{\caption{(a): The BPS quiver gauging
procedure for joining the Gaiotto curves for two $SU\left(2\right)$
Gaiotto theories with $N_{f}=4$. On the right and left we have the
BPS quivers for these Gaiotto theories. In the centre we have introduced
a pure $SU\left(2\right)$ subquiver to add the gauge degrees of freedom.
(b): The resulting BPS quiver.
The corresponding Gaiotto curve is a sphere with six punctures; this
corresponds to an $SU\left(2\right)^{3}$ theory with $N_{f}=6$.
This is constructed from the components in (a) by deleting
the nodes $g$ and $q$ and coupling $f$ and $p$ to the $SU\left(2\right)$
subquiver in oriented triangles.}}

\end{figure}

From this, one might be tempted to conclude that all $SU\left(2\right)$
Gaiotto theories are susceptible to such a translation between their
skeleton diagrams and BPS quivers. However, there are exceptions,
specifically for the case of $SU\left(2\right)$ theories with $g>2$
and $n=0$. Such theories admit no mass deformations 
(for $g<2$, one can see from the gluing procedure for skeleton diagrams
that we must always have at least one external leg, and thus must
have mass deformations), and hence do not admit BPS quivers \cite{CecottiVafa,FirstVafa}
- a point to which we shall return shortly in a different context.
The case of $g=2$ with no punctures is an exception: in one duality
frame, that theory corresponds to an $SU\left(2\right)^{3}$ theory
with two half-hypermultiplets; these two half-hypermultiplets form
one full hypermultiplet and that can receive mass \cite{CecottiVafa}.

Using these results, we can now write down a skeleton diagram (all
of which can be constructed by joining trivalent vertices) and immediately
compute a wealth of information about the associated $SU\left(2\right)$
Gaiotto theory:~the mutation class of BPS quivers, the BPS spectrum,
the ideal triangulations on the Gaiotto curve $\mathcal{C}$
corresponding to each BPS quiver, and the associated quadratic differentials
on $\mathcal{C}$ and Seiberg-Witten curves.

\subsection{Strebel Differentials and Ribbon Graphs}\label{s:ribbon}

\subsubsection{Ribbon Graphs from Strebel Differentials}

At a special point in the Coulomb branch $\mathcal{U}_{g,n}$ of the $SU\left(2\right)$ Gaiotto theory in question,
the coefficients of the quadratic differential will be such that it satisfies the definition of a so-called
\textbf{Strebel differential}. Choose an ordered $n$-tuple $\left(a_1,\ldots, a_n \right) \in \mathbb{R}^n_+$ of positive real numbers. Then, a Strebel
differential is a meromorphic quadratic differential $q$ on a Riemann
surface $\mathcal{C}$ of genus $g$ with $n$ marked points $\left\{ p_{1},\ldots,p_{n}\right\} $
(subject to the conditions $g\geq0$, $n\geq1$, and $2-2g-n<0$) satisfying \cite{Mulase}:
\begin{enumerate}
\item $q$ is holomorphic on $\mathcal{C}\setminus\left\{ p_{1},\ldots,p_{n}\right\} $.
\item $q$ has a second order pole at each $p_{j}$, $j=1,\ldots,n$.
\item The union of all non-compact horizontal trajectories forms a closed
subset of $\mathcal{C}$ of measure zero.
\item Every compact horizontal trajectory $\alpha$ is a simple loop circling
around one of the poles, say $p_{j}$, satisfying $a_{j}=\oint_{\alpha}\sqrt{q}$,
where the branch of the square root is chosen so that the integral
has a positive value with respect to the positive orientation of $\alpha$
that is determined by the complex structure of $\mathcal{C}$.
\end{enumerate}

Strebel differentials arise at specific points in $\mathcal{U}_{g,n}$. However, even at such points, the $a_i$ are unfixed; such numbers are naturally associated with the residues of the $n$ poles (physically, they can be associated with masses, couplings and moduli of the global flavour symmetries of the theory). Thus, to \emph{completely} fix a Strebel differential, we must fix a point in  $\mathcal{U}_{g,n} \times \mathbb{R}^n_+$.
Note that the condition $n\geq1$ ensures that no Strebel differential
can be defined on $g\geq2$, $n=0$ Gaiotto curves.
At the particular point in the Coulomb branch $\mathcal{U}_{g,n}$ of the theory where the quadratic differential is Strebel, 
the graph resulting from
joining the zeroes of $q$ via the horizontal trajectories, with one
marked point for each face, is known as a \textbf{ribbon graph} \cite{Mulase}.

\subsubsection{Ribbon Graphs as Dessins}

Now, it is at this point that Grothendieck's dessins d'enfants, i.e.~bipartite graphs drawn on Riemann surfaces, enter the story. First, fix the point $u \in \mathcal{U}_{g,n}$ to be a point where the quadratic differential becomes Strebel. At such a point, one can construct a ribbon graph on $\mathcal{C}$. Then, fix the $a_i$ to be such that the \emph{lengths} of the edges of the ribbon graph are all equal. To do this, recall from \cite{Mulase} that the lengths $L\left(E_i\right)$ of the $m$ edges of a ribbon graph enclosing its $k$th marked point are related to the associated number $a_k$ by $a_k= \sum_{i=1}^{m} L\left(E_i\right)$; writing down such an equation for all marked points, one can solve the system of simultaneous equations to determine the lengths of the edges of the ribbon graph. By convention, choose all these lengths to equal unity.

As discussed in \cite{Mulase}, at such a point in $\mathcal{U}_{g,n}\times \mathbb{R}^n_+$, the ribbon graph on $\mathcal{C}$ for each Gaiotto theory can be 
 can be interpreted as a clean dessin d'enfant, by colouring every vertex white and inserting a black
 node into every edge. In turn, we can associate a unique Belyi map to every such dessin by following the procedure
 detailed in \S2.5. The Belyi map $\beta\left(x\right)$ determined by the ribbon graph
  in this way then possesses an interesting
 property \cite{Mulase}: the Strebel differential $\phi\left(x\right)\mathrm{d}x^{2}$ on $\mathcal{C}$, fixing the $a_i$ so that all ribbon graph edges are equal to unity, can be
constructed as the pullback by $\beta\left(x\right)$ of a quadratic differential on $\mathbb{P}^1$ with
coordinates $\zeta$ with three punctures:

\begin{equation}
q=\phi\left(x\right)\mathrm{d}x^{2}=\beta^{*}\left(\frac{1}{4\pi^{2}}\frac{\mathrm{d}\zeta^{2}}{\zeta\left(1-\zeta\right)}\right).
\end{equation}

\noindent In other words, the map $\beta\left(x\right)$ from $\mathcal{C}$
to $\mathbb{P}^{1}$ is precisely the Belyi map corresponding to the
ribbon graph interpreted as a dessin \cite{Mulase}.
Clearly, it follows immediately from the definition of the pullback that we
can write:

\begin{equation}
q=\frac{1}{4\pi^{2}}\frac{\mathrm{d}\beta^{2}}{\beta\left(1-\beta\right)}.
\end{equation}

\noindent In itself, this is
an extremely intriguing result. However, we can go further, by now
recalling \textbf{Belyi's theorem}. This states that a non-singular
Riemann surface $\mathcal{C}$ has the structure of an algebraic curve
defined on $\overline{\mathbb{Q}}$ if and only if there is a Belyi
map from $\mathcal{C}$ onto $\mathbb{P}^{1}$ \cite{Mulase,DessinBook}.
By this theorem, our Gaiotto curves (save the exceptional $g\geq2$,
$n=0$ cases discussed, where we cannot drawn a ribbon graph on $\mathcal{C}$)
have the structure of algebraic curves defined on $\overline{\mathbb{Q}}$, at these particular points in $\mathcal{U}_{g,n} \times \mathbb{R}^n_+$.

\subsubsection{Strebel Differentials from Belyi Maps}

Suppose we are given one possible ribbon graph on $\mathcal{C}$. With
this in hand, (3.8) provides a means of 
directly computing the associated Strebel differential (for the point in $\mathbb{R}^n_+$ where all edges are equal):~we simply substitute the associated Belyi
map into this formula. For example, consider the ribbon graph given in figure 4(a), for the $SU\left(2\right)$,
$N_f=4$ theory. The Belyi map associated to this tetrahedral ribbon graph, as a dessin, is given in
\cite{Goins}:


\begin{equation}
\beta_\mathrm{tetra}\left(z\right)=-64\frac{z^3\left(z^3-1\right)^3}{\left(1+8z^3\right)^3}.
\end{equation}

\noindent The pre-images of $\left\{0,1,\infty\right\}$ for this Belyi map are, respectively, $\left\{3^4,2^6,3^4\right\}$ \cite{YMR}. Substituting $\beta_\mathrm{tetra}\left(z\right)$ into (3.8) gives:

\begin{equation}
q= -\frac{576 z \left(z^3-1\right)}{4\pi^2 \left(1+8 z^3\right)^2}.
\end{equation}

\noindent This quadratic differential has poles at $\left\{ -\frac{1}{2}, \frac{1}{2} \left(-1\right)^{1/3}, -\frac{1}{2} \left(-1\right)^{2/3}, \infty \right\}$. This is the Strebel differential on $\mathcal{C}$ associated to the tetrahedral ribbon graph (with equal length edges) for the
$SU\left(2\right)$, $N_f=4$ Gaiotto theory. Note that this correctly matches the generic expected form of 
the quadratic differential for the $SU\left(2\right)$, $N_f=4$ theory, as presented in \cite{Gaiotto, Dummies} and discussed in the following section of this paper.
Clearly, the above method provides an efficient means of computing explicit Strebel differentials on $\mathcal{C}$.

\subsubsection{Enumerating Ribbon Graphs}

At this point, a further question naturally arises:~for a given $SU\left(2\right)$ Gaiotto theory, how
do we enumerate all topologically distinct possible ribbon graphs? In order
to answer this question, we first need to know, for a given $SU\left(2\right)$ Gaiotto theory, the most general possible
form of the quadratic differential on $\mathcal{C}$. This can be computed in the following way. 
First, suppose we have a Gaiotto theory with $n$ punctures on $\mathcal{C}$, so that the associated ribbon graphs 
 have $n$ faces. Given this, the number of \emph{vertices} of the ribbon graphs can be computed using
Euler's formula, which relates the number of vertices $V$, edges
$E$ and faces $F$ of a graph drawn on a surface of genus $g$:

\begin{equation}
V-E+F=2-2g .
\end{equation}

\noindent Since we assume all our ribbon graphs have simple zeroes and are thus
trivalent and connected, we have $E=\frac{3}{2}V$. From the above
reasoning, $F=n$. Thus:

\begin{equation}
V=2n-4+4g.
\end{equation}

\noindent Thus we can construct the quadratic differential for the theory in
question given only the topology $\left\langle g,n\right\rangle $
of the skeleton: generically, this will have $n$ second-order poles and $2n-4+4g$ faces.
Note, however, that we can \emph{only }apply this
method for the class of $SU\left(2\right)$ Gaiotto theories which
admit such a differential; for $g\geq2$, $n=0$, this is not possible,
since in this case one cannot construct a quadratic differential on
$\mathcal{C}$ with only second order poles \cite{Mulase}, as we have already observed.
 Setting aside the exceptional case of $g=2$, $n=0$
already discussed, this result is reassuring. This is because we have
already found that for $g\geq2$, $n=0$, the theory in question does
not admit a BPS quiver. But since we can translate between the quadratic
differential and the BPS quiver as described, we would expect to find
that in these cases we cannot write down a quadratic differential
of the generic form described above. Indeed we now see this to be the case.

So, the generic form of a ribbon graph on $\mathcal{C}$ will have $n$ faces (corresponding to second-order poles of $q$), and 
$2n-4+4g$ vertices (corresponding to zeroes of $q$). Any ribbon graph which fulfils these topological criteria is a possible ribbon graph
for the Gaiotto theory in question, and its associated Strebel differential, for the case of equal length edges, can be computed via the ribbon graph's
corresponding Belyi map in the manner detailed above. For details on algorithmic procedures for enumerating
all possible trivalent graphs with a given number of vertices and edges, the reader is referred to the classic works \cite{G1,G2,G3}, as well as the discussion in \cite{HeJohn}.

\subsubsection{Connections to Modularity}

It is interesting to note that the six topologically distinct possible ribbon graphs for the $SU\left(2\right)$,
$N_f=4$ theory correspond to the dessins for the index 12 modular subgroups presented in 
\cite{MS, HeJohn, YMR}. For ease of reference, these are presented in Figure 6. 
With this in mind, the question arises as to which $SU\left(2\right)$ Gaiotto theories the remaining
dessins in \cite{YMR} correspond, insofar as they are possible ribbon graphs on $\mathcal{C}$.
 To answer this question, first recall that all the dessins in question are drawn on
the sphere, so we must have $g=0$. In addition, each dessin must have $n$ faces, where $n$ is the number
of punctures on $\mathcal{C}$ for the theory in question. From the work in the previous section, the dessin must therefore
have $V=2n-4$ vertices. 


\begin{figure}

\begin{center}

\begin{minipage}[t]{0.25\textwidth}%
\begin{center}
\includegraphics[scale=0.15]{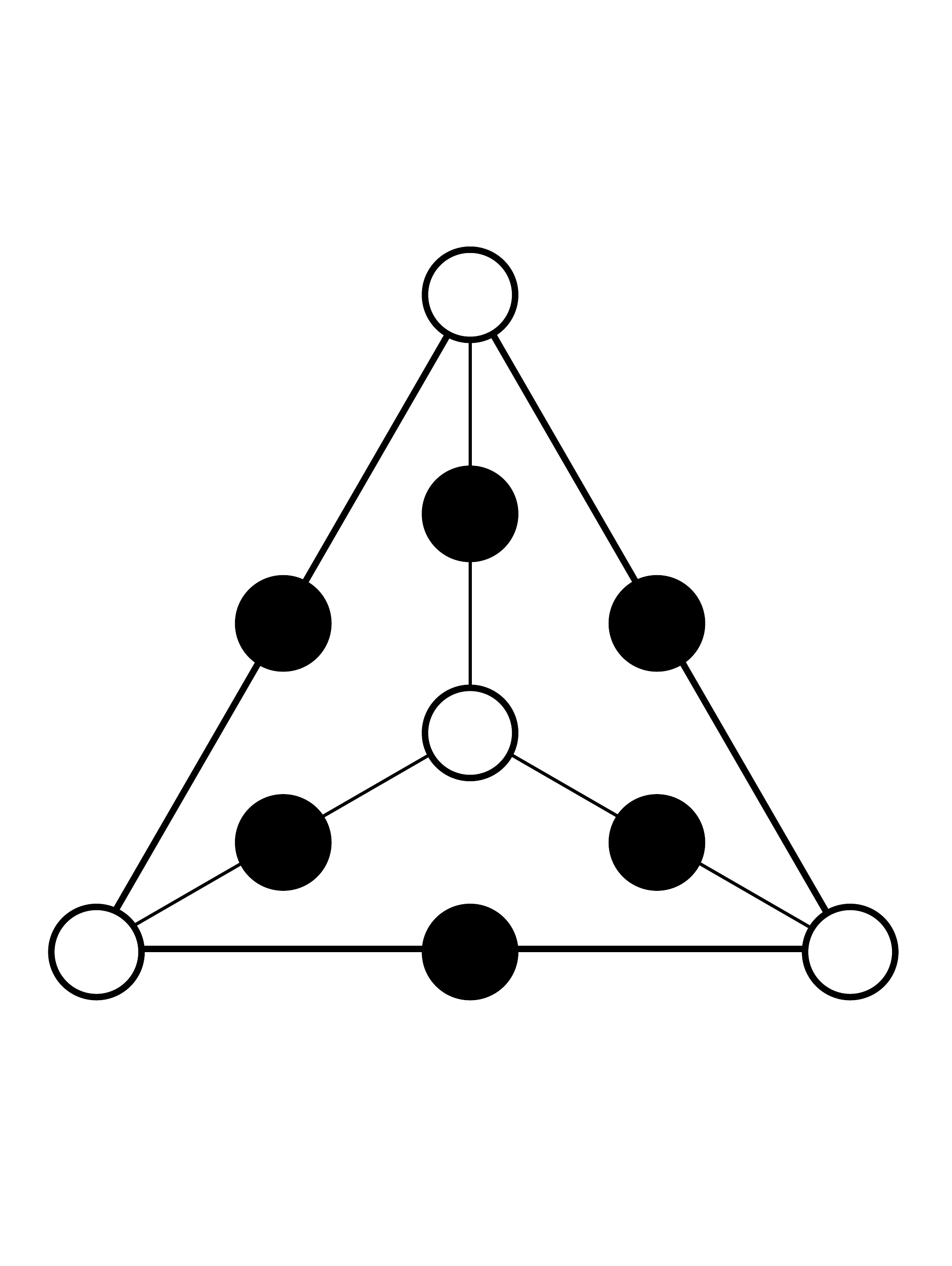}
\par
\end{center}
\end{minipage}\qquad{}%
\begin{minipage}[t]{0.25\textwidth}%
\begin{center}
\includegraphics[scale=0.15]{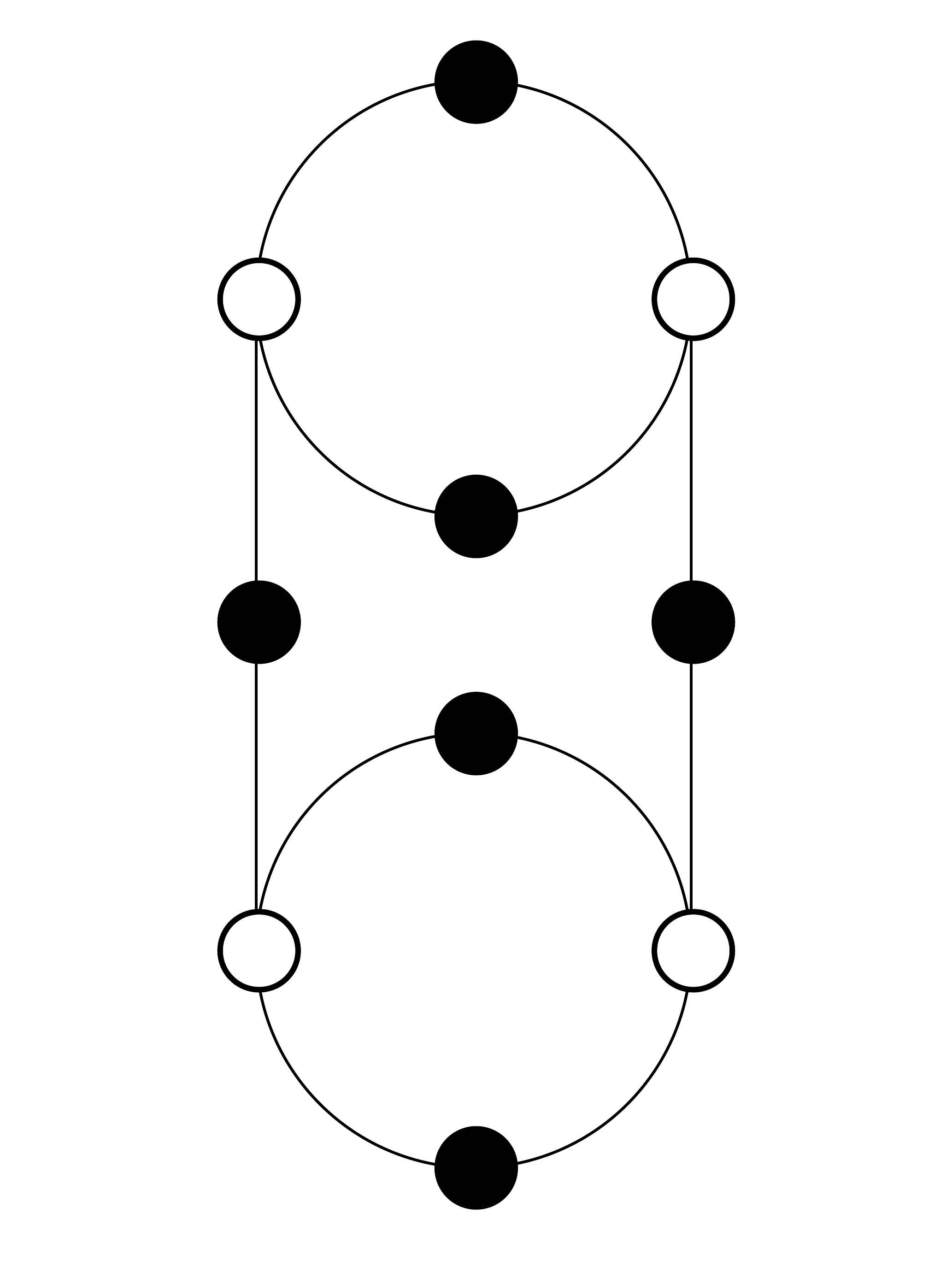}
\par
\end{center}
\end{minipage}\qquad{}%
\begin{minipage}[t]{0.25\textwidth}%
\begin{center}
\includegraphics[scale=0.15]{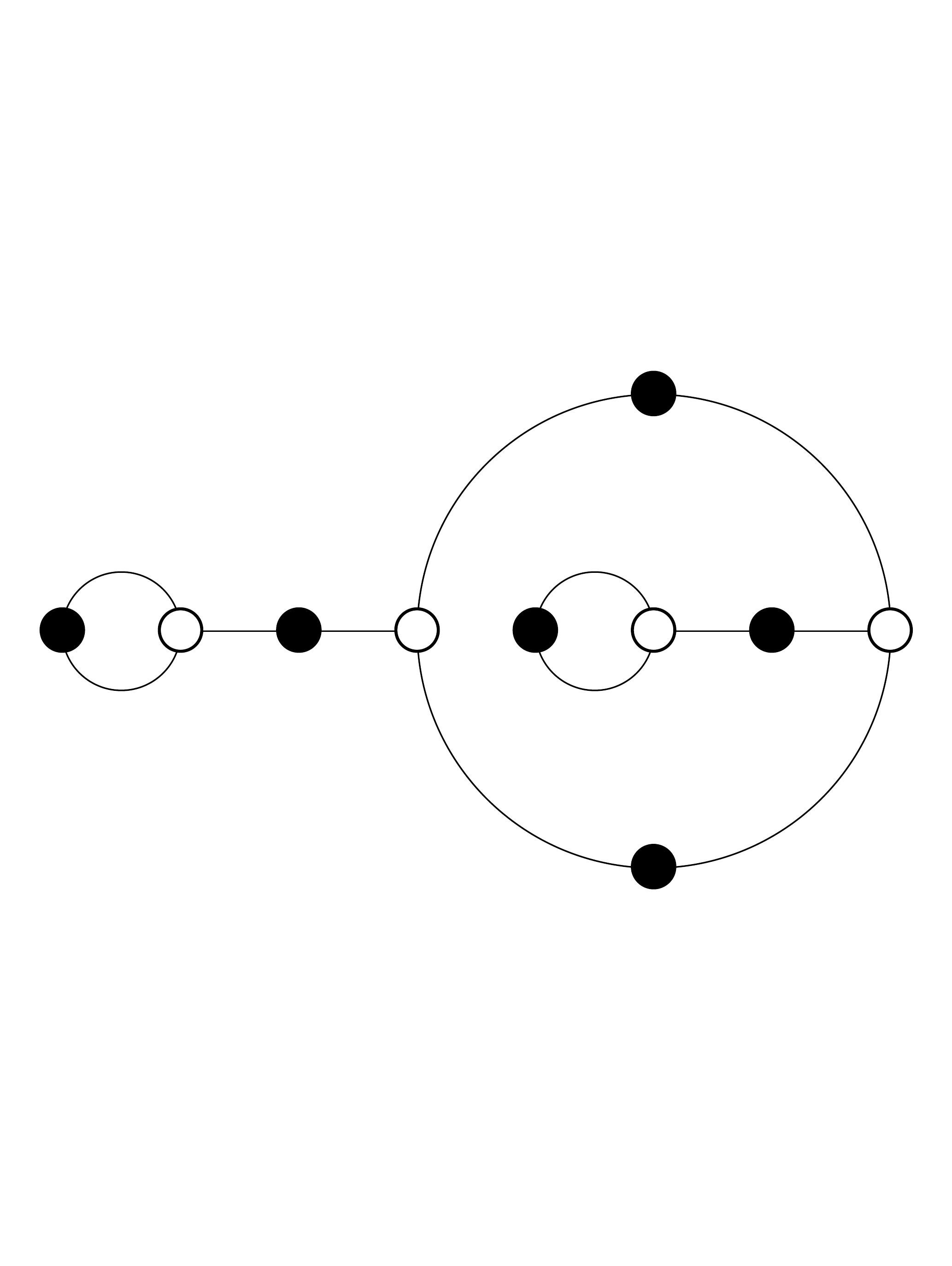}
\par
\end{center}
\end{minipage}\qquad{}%
\begin{minipage}[t]{0.25\textwidth}%
\begin{center}
\includegraphics[scale=0.15]{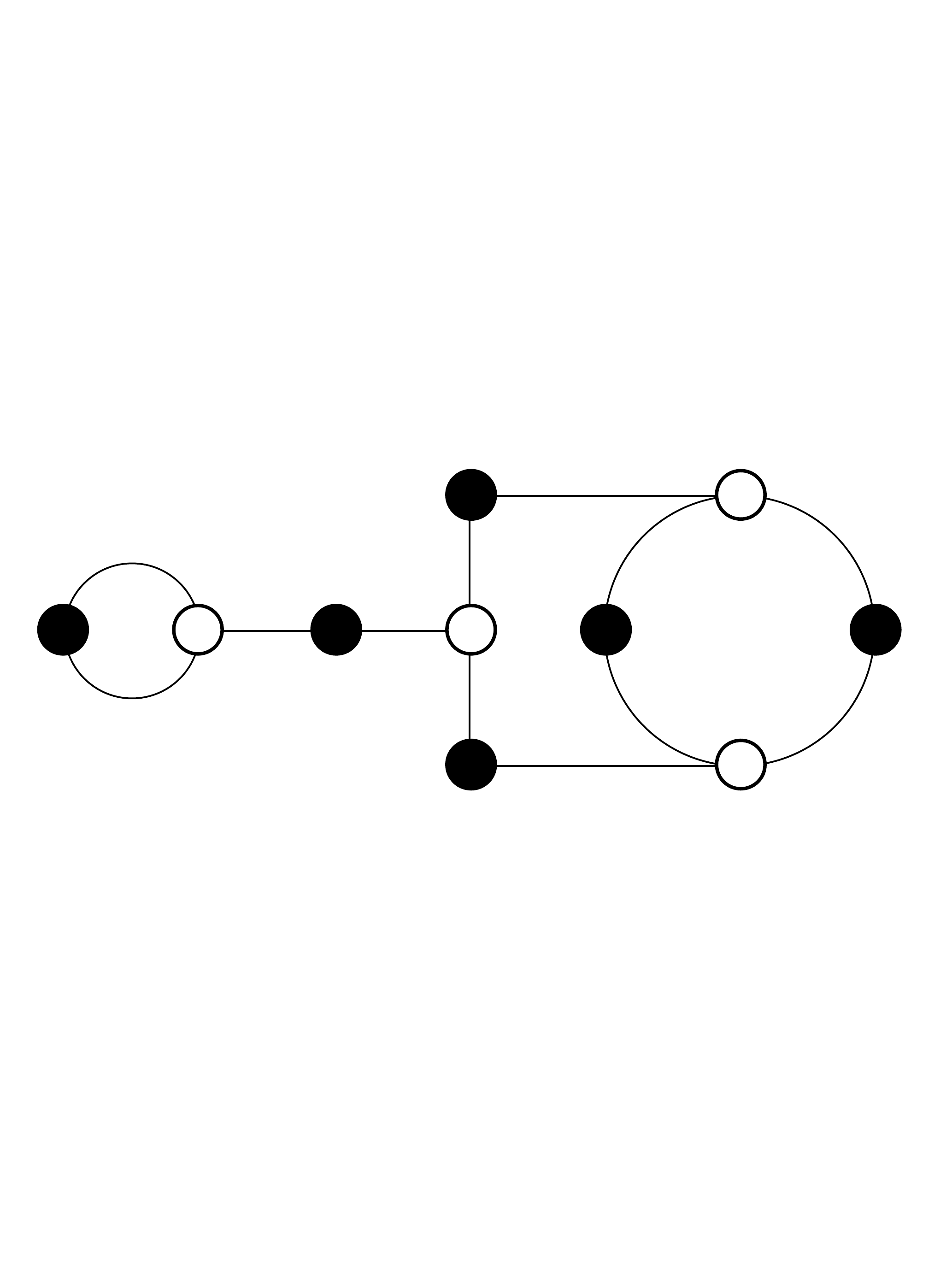}
\par
\end{center}
\end{minipage}\qquad{}%
\begin{minipage}[t]{0.25\textwidth}%
\begin{center}
\includegraphics[scale=0.15]{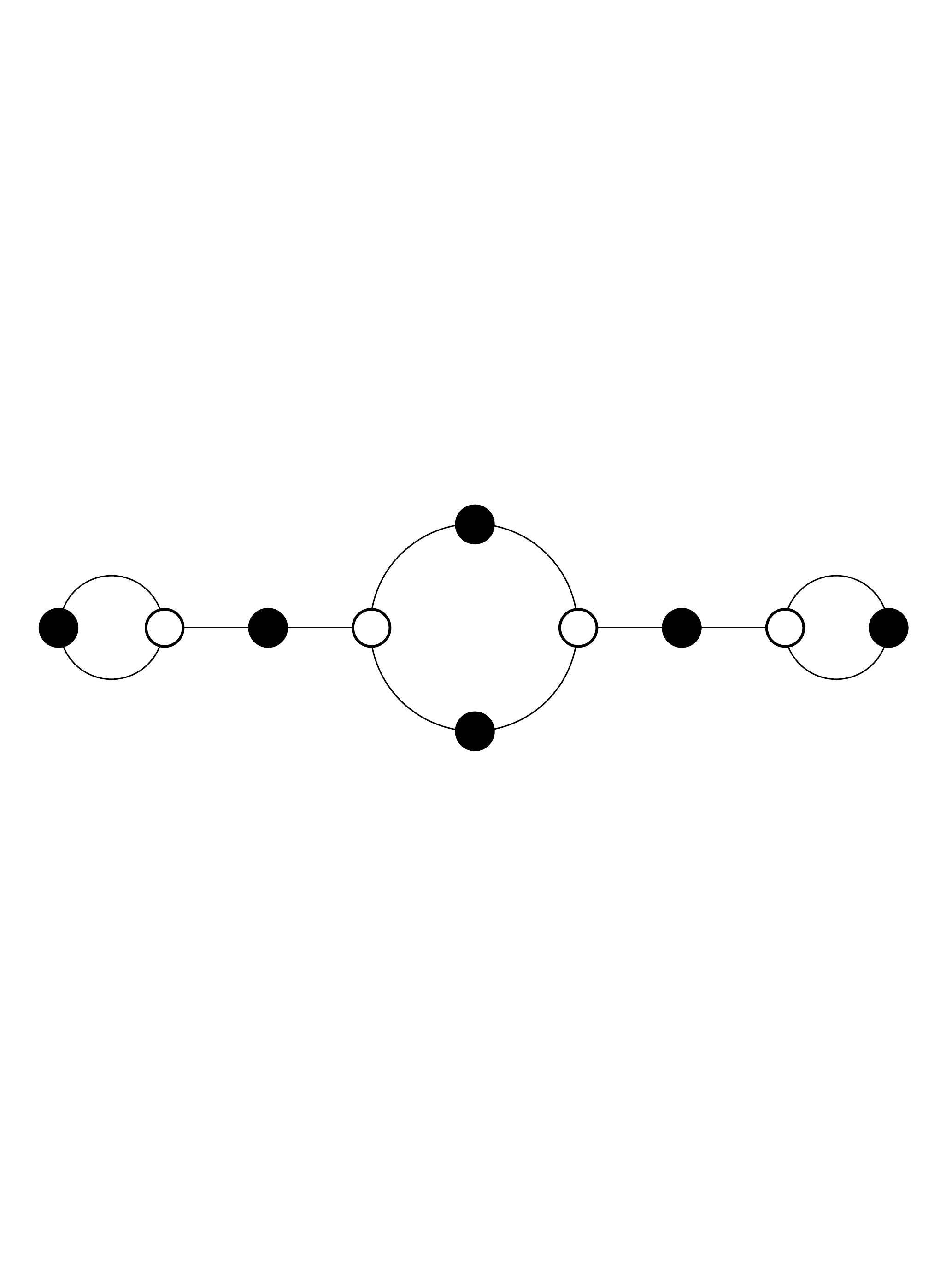}
\par
\end{center}
\end{minipage}\qquad{}%
\begin{minipage}[t]{0.25\textwidth}%
\begin{center}
\includegraphics[scale=0.15]{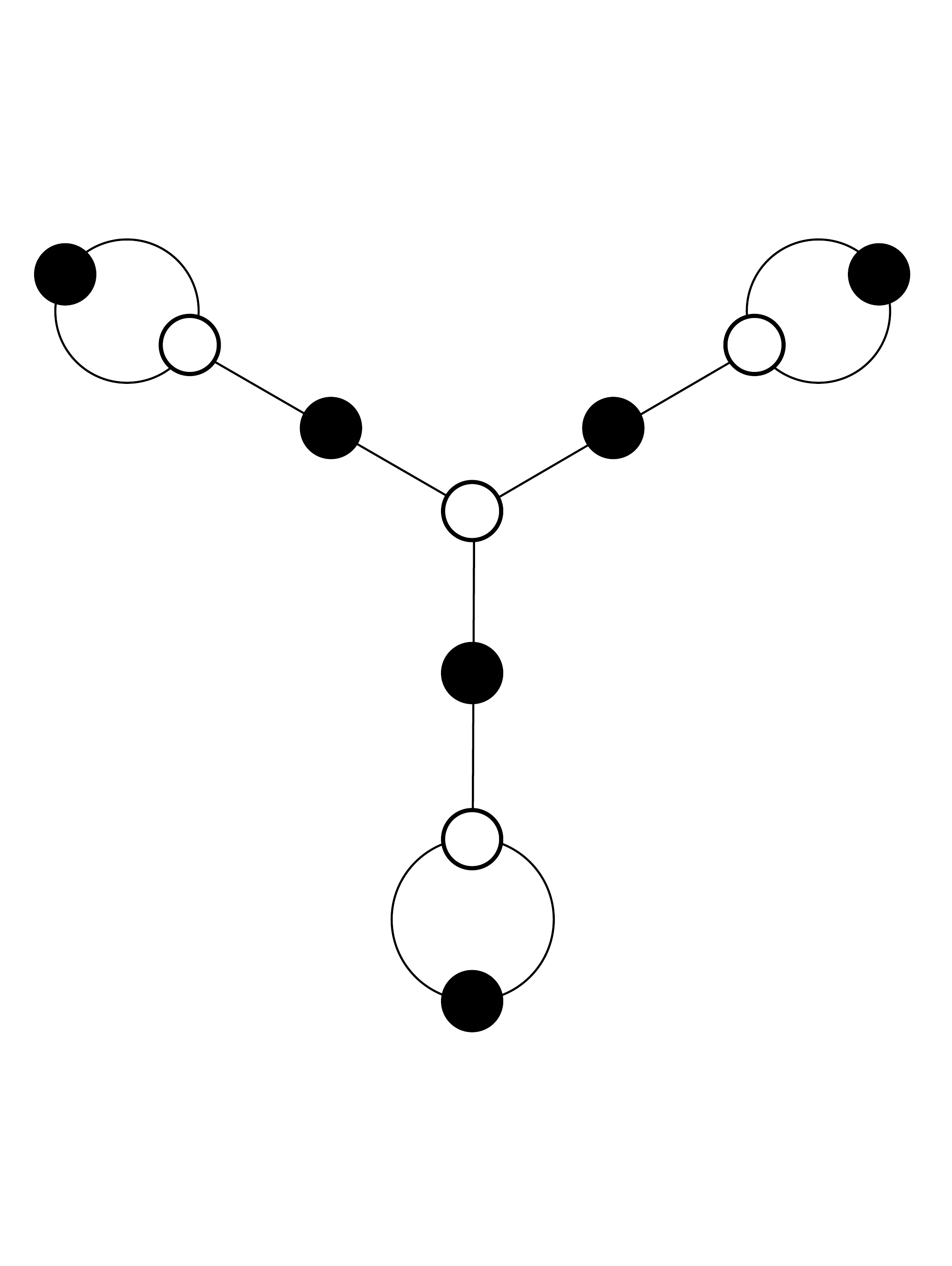}
\par
\end{center}
\end{minipage}

%
\par\end{center}

\sf{\caption{The six possible ribbon graph topologies for the 
$SU\left(2\right)$, $N_{f}=4$ theory, drawn as dessins d'enfants. These are precisely the six
index 12 dessins in \cite{YMR}.}}

\end{figure}


\begin{table}

\begin{center}
\begin{minipage}[t]{1\textwidth}%
\noindent \begin{center}
\begin{tabular}{|c|c|c|c|}
\hline 
Number of faces $\left(n\right)$ & Number of vertices $\left(2n-4\right)$ & Index of dessin & Gaiotto theory\tabularnewline
\hline 
\hline 
4 & 4 & 12 & $SU\left(2\right)$, $N_f=4$\tabularnewline
\hline 
6 & 8 & 24 & $SU\left(2\right)^3$, $N_f=6$\tabularnewline
\hline 
8 & 12 & 36 & $SU\left(2\right)^5$, $N_f=8$\tabularnewline
\hline 
10 & 16 & 48 & $SU\left(2\right)^7$, $N_f=10$\tabularnewline
\hline 
12 & 20 & 60 & $SU\left(2\right)^9$, $N_f=12$\tabularnewline
\hline 
\end{tabular}
\par\end{center}

\end{minipage}
\par\end{center}

\sf{\caption{The number of faces and vertices of a ribbon graph drawn on a genus zero
Gaiotto curve, as well as the index of the dessins from \cite{YMR} which match those numbers of faces and
vertices.}}
\end{table}

The number of faces and vertices for important values of $n$ are tabulated in Table 1.
The reader can note that the number of faces and vertices of a ribbon graph drawn on a genus zero
Gaiotto curve precisely match those of the various index dessins in \cite{YMR}, as given in the third column of Table 1.
Hence, one can see that the dessins in \cite{YMR} \emph{do} correspond to ribbon graphs of certain Gaiotto theories.

Which Gaiotto theories are these? Since the skeleton diagrams for an $SU\left(2\right)$ Gaiotto
theory can all be constructed by joining trivalent vertices, for a genus zero Gaiotto curve, we simply need to join
repeated trivalent vertices, without loops, until the right number of punctures is reached. (Recall that each internal leg of a skeleton diagram corresponds to an $SU\left(2\right)$
gauge group factor; each external leg corresponds to a puncture, and thereby to an external flavour symmetry.)
The resulting skeleton diagram with $n$ external legs gives the Gaiotto theory
to which the dessin in question -- with $n$ vertices and $2n-4$ faces -- corresponds.
The results of undertaking this process are given in the fourth and final column of Table 1.

\subsubsection{Location of the Strebel Points in the Coulomb Branch}

What, then, is the physical significance of these ribbon graphs, which arise where the quadratic differential
on $\mathcal{C}$ satisfies the definition of a Strebel differential? As stated in \S6 of \cite{GMN}, flow
lines form closed orbits around marked points precisely where a BPS state appears and the topology of the 
triangulation (and thus BPS quiver) jumps. Hence, if the Coulomb branch is partitioned into domains for
each BPS quiver of the theory, these Strebel points in the moduli space must arise at the walls \emph{separating}
these domains.

\subsubsection{Dessins at Other Points in the Moduli Space}

It is worth making a further comment on the form (3.8) of the quadratic differential $q$ in terms of a Belyi map
$\beta$. Though at a Strebel point $q$ can be written in this form, with the Belyi map then being that associated
to the ribbon graph (with equal length edges) interpreted as a dessin, this does not preclude us from being able to write $q$ in the form (3.8) at some
\emph{other} isolated points in the moduli space.\footnote{We thank Diego Rodriguez-Gomez for pointing out this possibility, and for the calculations which follow in this subsection.} The reason for this is that Theorem 6.5 of \cite{Mulase} is an `if' rather
than an `if and only if' statement in this respect.
Indeed, we can find such a point in the moduli space as in the following example. Consider again the $SU\left(2\right)$, $N_f=4$ theory, but this time
begin with the generic form of the quadratic differential, with four simple zeroes and four second-order poles. We parameterise this generic form as

\begin{equation}
\phi\left(x\right) = \frac{\left(u_1 x^2 +m_1 x + l_1 \right)^2 - 4\left(u_0 x^2 + m_0 x + l_0 \right) \left(u_2 x^2 + m_2 x + l_2 \right)}{4x^2\left(u_2 x^2 + m_2 x + l_2\right)^2},
\end{equation}

\noindent with $q=\phi\left(x\right)\mathrm{d}x^2$. Now, choosing

$$
u_2 = - \left( 4+ 3\sqrt{2}\right)\pi \qquad u_1 = i\left(4+3\sqrt{2}\right) - \frac{\sqrt{2}}{3}m_1 \qquad u_0 = \frac{6i\sqrt{2}+\left(4-3\sqrt{2}\right)m_1}{36 \pi} m_1
$$
$$
l_2 = - \left(4+3\sqrt{2}\right)\pi \qquad l_1 = \frac{3i\sqrt{2} + \left(3\sqrt{2}-4\right)}{6\sqrt{2} - 9}m_1 \qquad l_0 = - \frac{6i\sqrt{2} + \left(3\sqrt{2}-4\right)m_1}{36\pi}m_1
$$
\begin{equation}
m_2 = 3\left(3+2\sqrt{2}\right)\pi \qquad m_0 = \frac{36+24\sqrt{2} +\left(3-2\sqrt{2}\right)m_1^2}{12\pi}
\end{equation}

\noindent we have, upon doing the transformation

\begin{equation}
x \rightarrow \frac{x-\frac{1}{2}\left(1-i\sqrt{3-2\sqrt{2}}\right)}{x-\frac{1}{2}\left(1+i\sqrt{3-2\sqrt{2}}\right)}
\end{equation}

\noindent a differential in our ``canonical'' Strebel form (3.8), this time with Belyi map

\begin{equation}
\beta\left(x\right) = \frac{x^4}{x^4+\left(x-1\right)^4}.
\end{equation}

\noindent What is the dessin corresponding to this Belyi map? In fact, it is known from \cite{Goins, Diego} that a Belyi map of the 
form

\begin{equation}
\beta_N\left(x\right) = \frac{x^N}{x^N+\left(x-1\right)^N}
\end{equation}

\noindent has an associated dessin of the form shown in Figure (8).
\begin{figure}

\begin{center}

\begin{minipage}[t]{0.25\textwidth}%
\begin{center}
\includegraphics{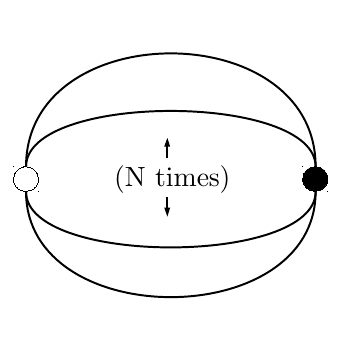}
\par
\end{center}
\end{minipage}\qquad{}%

\par\end{center}

\sf{\caption{Dessin for the Belyi map $\beta_N\left(x\right)$. For $N=4$, we have a 4-valent white node joined to
a 4-valent black node on $\mathbb{P}^1$.}}

\end{figure}


The first thing to notice about this dessin is that it is \emph{not} {clean}. But since all the dessins associated to ribbon graphs
are by construction clean, this means that this dessin cannot correspond to a ribbon graph. In turn, this means 
that this specific point in the moduli space cannot be Strebel -- i.e.~$q$ is not a Strebel differential at this point. Hence we see
that, starting from the generic expression for the quadratic differential for a certain theory and tuning parameters in
the way describe above, one does not \emph{necessarily} arrive at a form (3.8) which corresponds to a Strebel point
(note, though, that all the Strebel points can be found simply by fixing parameters in $q$, whereas in the 
above we also needed to transform $x$).

Whether these non-Strebel dessins have any significance is an open question. One must approach such results with a certain degree of caution,
since it is not clear what significance the form of the quadratic differential (3.8) has away from the Strebel points.
Nevertheless, these auxiliary dessins which arise at other points in the moduli space in this
way present an interesting opportunity for future investigation.

\subsubsection{Further Conjectures}

Based on the above work, one might attempt to link previous work on the connections between dessins
d'enfants and $\mathcal{N}=2$ $U\left(N\right)$ gauge theories presented in
\cite{FirstCachazo,CachazoDessins} to the $SU\left(2\right)$ Gaiotto case.
In this section we will see, however, that the analogy is at least not a direct one, and several
aspects of it fail.

To begin, first recall that in \cite{FirstCachazo} the authors demonstrate how the problem of finding Argyles-Douglas
singularities in the Coulomb branch $\mathcal{U}$ of an $\mathcal{N}=2$ theory with $U\left(N\right)$ gauge group
 can be mapped to the problem of finding when an abstractly defined quadratic differential on a Riemann
 surface becomes Strebel. Moreover, at these special Argyres-Douglas points, the Belyi map associated
 to the ribbon graph (interpreted as a dessin) for that Strebel differential
 can be used to construct the Seiberg-Witten curve of the theory.
  This is a purely formal correspondence, 
but the work above {suggests} that these quadratic differentials and dessins have a nice interpretation
for $SU\left(2\right)$ Gaiotto theories:~the quadratic differentials are precisely the
quadratic differentials on the Gaiotto curves which appear in the Seiberg-Witten curves,
while the dessins d'enfants are precisely the ribbon graphs drawn on the Gaiotto curves.

Is this conjecture correct? To evaluate it, we must recall some further details from \cite{FirstCachazo}. In that
paper, the authors consider the ribbon graph associated to the abstractly defined quadratic differentials.
If we take such a ribbon graph and interpret as a dessin, we can find the associated Belyi map, which can generically
be expressed as

\begin{equation}
\beta\left(z\right) = \frac{A\left(z\right)}{B\left(z\right)},
\end{equation}

\noindent where $A\left(z\right)$ and $B\left(z\right)$ are polynomials of some degree.
In turn, this Belyi map can be used to construct the Seiberg-Witten curve for the $U\left(N\right)$ gauge
theory in question via the identification

\begin{equation}
y^2 = P^2\left(z\right)+B\left(z\right),
\end{equation}

\noindent where

\begin{equation}
P^2\left(z\right) = A\left(z\right) - B\left(z\right).
\end{equation}

Given this identification of the  polynomials of the Belyi map with the right hand side of the Seiberg-Witten curve
in hyperelliptic form, it 
is clear that if the quadratic differential discussed in \cite{FirstCachazo} can indeed be interpreted as the 
quadric differential on the Gaiotto curve in the case of $SU\left(2\right)$ Gaiotto theories,
 it must be the case that at the points in the moduli space
of the theory where this becomes Strebel, the associated Belyi map yields a Seiberg-Witten curve
in hyperelliptic form of the correct degree for the theory in question.
This is a proposition which can be easily tested in a concrete example;
 we will choose for simplicity the $SU\left(2\right)$, $N_f=4$ theory.

The Belyi map for the tetrahedral ribbon graph for this theory is given in (3.10). From this, we can see that
the numerator is a degree 9 polynomial in $z$. Hence, on the above prescription, the Seiberg-Witten curve
for this theory has the form $y^2 = A_9\left(z\right)$. However, the Seiberg-Witten curve
for this theory is in fact of degree \emph{four}. We can reason to this answer in the following way.


\begin{figure}

\begin{center}

\begin{minipage}[t]{0.25\textwidth}%
\begin{center}
\includegraphics{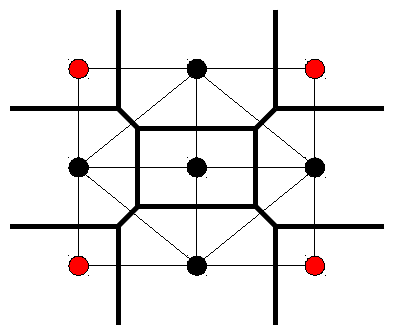}
\par
\end{center}
\end{minipage}\qquad{}%

\par\end{center}

\sf{\caption{Braneweb and grid diagrams for the $SU\left(2\right)$, $N_f=4$ theory.}}

\end{figure}

First, this theory can be seen as 
the dimensional reduction of the five dimensional theory living on the braneweb shown in Figure 8.
The Seiberg-Witten curve corresponding to the 5D theory living on the sphere is \cite{HananyKol}

\begin{equation}
\tilde{L}_2 + \tilde{y}\tilde{M}_2 + \tilde{y}^2\tilde{U}_2 = 0 ,
\end{equation}

\noindent where $\tilde{L}_2$, $\tilde{M}_2$ and $\tilde{U}_2$ are degree two polynomials in $\tilde{x}$ whose
coefficients are associated to the dots in the grid diagram. In this curve, the holomorphic two-form is
$\mathrm{d}\lambda = \mathrm{d}\log x \wedge \mathrm{d}\log y$. The standard reduction to 4D amounts to
taking $\tilde{y}=y$ and $\tilde{x}=e^{2\epsilon x}$. In the appropriate $\epsilon\rightarrow 0$ limit \cite{Brand, Wit},
the curve becomes

\begin{equation}
{L}_2 + {y}{M}_2 + {y}^2{U}_2 = 0 ,
\end{equation}

\noindent with the expected holomorphic two-form $\mathrm{d}x\wedge \mathrm{d}\log y$. Note that the polynomials
$L_2$, $M_2$ and $U_2$ are not the same as $\tilde{L}_2$, $\tilde{M}_2$ and $\tilde{U}_2$. Indeed,
the first terms in the $\epsilon$ expansion of the coefficients of the latter reshuffle in some way to construct
the former. Now, upon doing

\begin{equation}
y=\frac{1}{U_2}\left(t-\frac{M_2}{2}\right) ,
\end{equation}

\noindent we find:

\begin{equation}
t^2 = \frac{M_2^2}{4} - L_2 U_2 .
\end{equation}

\noindent Clearly, the hyperelliptic curve for this theory is degree four in the right hand side, not degree nine.
The upshot of this is that we cannot use the Belyi map associated to a given Strebel differential on $\mathcal{C}$ to
construct the appropriate Seiberg-Witten curve, as for the abstractly defined quadratic differentials in 
\cite{FirstCachazo,CachazoDessins}. Of course, it is still possible that the points
in the Coulomb branch at which the quadratic differential on $\mathcal{C}$ becomes Strebel give interesting
factorisations of the Seiberg-Witten curves as in \cite{FirstCachazo,CachazoDessins}, but more work needs to be done
to establish this point.
At the very least, the connections to the work of the cited papers is not as straightforward as one might hope.

\subsection{Taking Stock}

Let us briefly recap the results gathered up to this point. Take a Gaiotto curve $\mathcal{C}$ of genus $g$
with $n$ punctures. The Seiberg-Witten curve for this theory has the form $y^2=\phi\left(x\right)$, where
$q=\phi\left(x\right)\mathrm{d}x^2$ is a meromorphic quadratic differential on $\mathcal{C}$.
The precise number of zeroes and second-order poles which this quadratic differential possesses was computed
in \S3.5.4. From the quadratic differential one can construct an ideal triangulation, and in turn the mutation
class of BPS quivers for the theory in question, as detailed in \S3.2-3.4.

The parameters of the quadratic differential vary as one varies the point in the Coulomb branch $\mathcal{U}_{g,n}$
under consideration. At certain very special points, $q$ will satisfy the definition of a Strebel differential; at these
points we can draw a ribbon graph on $\mathcal{C}$. To completely fix $q$, we must fix $n$ further positive real parameters, associated to the lengths of the edges of the ribbon graph. Fixing these parameters such that the edge lengths are unity (and thus completely fixing $q$ by fixing a point in $\mathcal{U}_{g,n} \times \mathbb{R}^n_+$), this ribbon graph can be interpreted as a trivalent
dessin d'enfant, with an associated Belyi map. The Belyi map relates $q$ at this point to a meromorphic quadratic differential on 
$\mathbb{P}^1$ by pullback, as detailed in \S3.5.1-3.5.2. In this way, as detailed in \S3.5.3, we can
reconstruct Strebel differentials at such points just given possible ribbon graph topologies on $\mathcal{C}$.

In \S3.5.5, we identified the Gaiotto theories to which the dessins in \cite{YMR} correspond, insofar
as they are possible ribbon graphs for those theories; in \S3.5.6 we identified the location of the Strebel
points in the Coulomb branch of these theories.
In \S3.5.7 we
investigated the possibility of the form (3.8) of the quadratic differential arising at non-Strebel points.
Finally, in \S3.5.8, we demonstrated that one cannot
straightforwardly identify the quadratic differential on $\mathcal{C}$ with the abstractly defined quadratic differentials
in \cite{FirstCachazo, CachazoDessins}, as doing so yields inconsistent results. 


\section{Skeleton Diagrams to Seiberg-Witten Curves: An Alternative Route?}

In \cite{HeJohn}, it is stated that the skeleton diagrams should
be interpreted \emph{directly} as dessins d'enfants; from there it
is claimed that we can construct the corresponding Seiberg-Witten
curve by manipulating the Belyi map associated to this dessin. In this section we show that this
deployment of dessins cannot work in general, as the method cannot guarantee that
the Seiberg-Witten curve will have the correct form.
To do this, we follow the methodology of \cite{HeJohn}, where the authors consider the specific 
class of dessins corresponding to the 33 {genus zero, torsion-free congruence
subgroups} of the modular group $\Gamma$ (introduced in \S2.6), all of which have $g>0$ and $n=0$, interpreting these
as skeleton diagrams.

The general setup is as follows: we suppose that we have a skeleton diagram with $g$ loops and
$n$ external legs, topologically identical to one of dessins in \cite{HeJohn}. We interpret this as a
dessin: exactly the corresponding dessin in \cite{HeJohn}. We then attempt to follow the prescription
in \cite{HeJohn} to construct the Seiberg-Witten curve corresponding to the original skeleton diagram
from the Belyi map for the associated dessin. We do this for a wide class of skeletons (all of which
are topologically identical to dessins in \cite{HeJohn}), showing that the proposed method
fails for some of these skeleton diagrams. Therefore, the proposed route from skeleton
diagrams to Seiberg-Witten curves needs to be modified, in the manner presented below.

With the above in mind, let us begin our investigations.
First recall from \cite{CecottiVafa} that the genus
of the Seiberg-Witten curve $g\left(\Sigma\right)$ is related to
the genus of the Gaiotto curve $g\left(\mathcal{C}\right)$ by the
following formula, where $p_{i}$ denotes a puncture on $\mathcal{C}$,
and we consider separately punctures of odd and even order:

\begin{equation}
g\left(\Sigma\right)=4g\left(\mathcal{C}\right)-3+\frac{1}{2}\sum_{p_{i}\,\mathrm{even}}p_{i}+\frac{1}{2}\sum_{p_{i}\,\mathrm{odd}}\left(p_{i}+1\right) .
\end{equation}
The genus of the Gaiotto curve $g\left(\mathcal{C}\right)$ is determined
by the number of loops of the skeleton diagram, so we find, by considering the 
number of loops of each of these dessins in \cite{YMR},
that for a skeleton diagram (interpreted as a dessin) corresponding
to a genus zero, torsion-free, congruence subgroup of index $I$,
the genus of the Gaiotto curve is given by

\begin{equation}
g\left(\mathcal{C}\right)=\frac{I}{6}+1 .
\end{equation}

\noindent Thus we have from (4.2) and (4.3) (noting that in our case all $p_i$ are even, since they are of order 2):

\begin{equation}
g\left(\Sigma\right)=\frac{2I}{3}+1 .
\end{equation}

Now, any hyperelliptic curve has the equation $y^{2}=Q_{N}\left(x\right)$,
where $Q_{N}\left(x\right)$ is a degree $N$ polynomial in $x$.
A genus $g\left(\Sigma\right)$ hyperelliptic curve has the equation
$y^{2}=Q_{2g\left(\Sigma\right)+1}\left(x\right)$ (for an imaginary
hyperelliptic curve) or $y^{2}=Q_{2g\left(\Sigma\right)+2}\left(x\right)$
(for a real hyperelliptic curve). Using our above result for $g\left(\mathcal{C}\right)$,
we therefore find that the Seiberg-Witten curves corresponding to
our index $I$ dessins can be written in the form $y^{2}=Q_{4I/3+3}\left(x\right)$
in the imaginary case, and $y^{2}=Q_{4\left(I/3+1\right)}\left(x\right)$
in the real case. Thus the degrees of the polynomials in the Seiberg-Witten
curves for the indices $I$ of interest from \cite{HeJohn} (i.e. $I\in\left\{ 6,12,24,36,48,60\right\} $)
are as shown in Table 2.

\begin{table}

\begin{center}
\begin{minipage}[t]{1\textwidth}%
\noindent \begin{center}
\begin{tabular}{|c|c|c|}
\hline 
Index $I$ & $\deg\left(Q_{4I/3+3}\left(x\right)\right)$ & $\deg\left(Q_{4\left(I/3+1\right)}\left(x\right)\right)$\tabularnewline
\hline 
\hline 
6 & 11 & 12\tabularnewline
\hline 
12 & 19 & 20\tabularnewline
\hline 
24 & 35 & 36\tabularnewline
\hline 
36 & 51 & 52\tabularnewline
\hline 
48 & 67 & 68\tabularnewline
\hline 
60 & 83 & 84\tabularnewline
\hline 
\end{tabular}
\par\end{center}

\end{minipage}
\par\end{center}

\sf{\caption{The degrees of the Seiberg-Witten
curves for an $\mathcal{N}=2$ $SU\left(2\right)$ Gaiotto theory
where the skeleton diagram is a dessin d'enfant corresponding to a
genus zero, torsion-free, congruence subgroup of index $I$. The second
column corresponds to the imaginary hyperelliptic case; the third
to the real hyperelliptic case.}}
\end{table}

For a dessin corresponding to an index $I$ subgroup, the corresponding
Belyi map $\beta\left(x\right)$ is a quotient of two polynomials
$A\left(x\right)$ and $B\left(x\right)$, the difference of which
is equal to the square of a polynomial of degree $I/2$ \cite{HeJohn,CachazoDessins}:

\begin{equation}
A\left(x\right)-B\left(x\right)=P_{I/2}^{2}\left(x\right) .
\end{equation}

It is at this point that the conjecture in this approach begins\emph{.}
We need some procedure taking us from $P_{I/2}\left(x\right)$ on
the dessin side to $Q_{4\left(I/3+1\right)}\left(x\right)$ on the
Seiberg-Witten side (focussing on the case of real hyperelliptic curves - the situation
for imaginary hyperelliptic curves is analogous). As a matter of simply matching degrees, we have:

\begin{equation}
a\cdot\deg\left(P_{I/2}\left(x\right)\right)=\deg\left(Q_{4\left(I/3+1\right)}\left(x\right)\right) .
\end{equation}

\noindent Solving for $a$, we find:

\begin{equation}
a=8\left(\frac{1}{3}+\frac{1}{I}\right) .
\end{equation}

\noindent For the indices of interest in \cite{HeJohn} ($I\in\left\{ 6,12,24,36,48,60\right\} $),
we have the results shown in Table 3.

\begin{table}

\begin{center}
\begin{minipage}[t]{1\textwidth}%
\noindent \begin{center}
\begin{tabular}{|c|c|}
\hline 
Index $I$ & Exponent $a$ in (4.6)\tabularnewline
\hline 
\hline 
6 & 4\tabularnewline
\hline 
12 & $10/3$\tabularnewline
\hline 
24 & 3\tabularnewline
\hline 
36 & $26/9$\tabularnewline
\hline 
48 & $17/6$\tabularnewline
\hline 
60 & $14/5$\tabularnewline
\hline 
\end{tabular}
\par\end{center}

\end{minipage}
\par\end{center}

\sf{\caption{The power $a$ to which a polynomial
of degree $I/2$ must be raised to produce a function (possibly polynomial)
of degree $4\left(I/3+1\right)$. This procedure allows us to match
the degree of $P_{I/2}\left(x\right)$ from the Belyi map corresponding
to the skeleton diagram interpreted as a dessin d'enfant to the expected
degree of the Seiberg-Witten curve for the Gaiotto theory in question.}}

\end{table}

The thought at this point is that we can then simply identify $P_{I/2}^{a}\left(x\right)$ from
the Belyi map associated to the skeleton diagram interpreted as a dessin with
$Q_{4\left(I/3+1\right)}\left(x\right)$:~the Seiberg-Witten curve for that gauge theory in 
hyperelliptic form.
Clearly, there is something special about index 6 and index 24, if
this procedure for going from $P_{I/2}\left(x\right)$ to $Q_{4\left(I/3+1\right)}\left(x\right)$
is correct. This is because only index 6 and index 24 give integer
$a$, and therefore guarantee polynomial $Q$, as required when constructing
a Seiberg-Witten curve. \emph{But for every $SU\left(2\right)$ gauge
theory of Gaiotto type, there is a Seiberg-Witten curve which can
be associated with the skeleton diagram.} Therefore, we see that,
when dessins for which $I\notin\left\{ 6,24\right\} $ are considered,
the fact that this method cannot guarantee that $Q\left(x\right)$
is a polynomial demonstrates that it is in general\emph{ not correct}.
Thus, in such cases, and hence in general, interpreting the skeleton
diagram as a dessin d'enfant and attempting to construct the corresponding
Seiberg-Witten curve from the Belyi map in this way will not work.

Indeed, there is no reason to suspect any direct connection between
skeleton diagrams and Seiberg-Witten curves via the theory of dessins
d'enfants (although, as, we have seen, dessins \emph{do} arise in
the context of the ribbon graphs). Moreover, this method clearly only
works when we consider skeleton diagrams without external legs, since
it is unclear what external legs of a dessin would mean. To conclude
then:~the correct method for matching the skeleton diagrams to the
corresponding quadratic differentials and Seiberg-Witten curves is
outlined in \S2 and \S3 of this paper; the work there supersedes the work presented in this section.


\section{Conclusions and Outlook}

In this paper, we have first recapitulated several significant results from 
\cite{MainVafa,CecottiVafa,FirstVafa,Hanany, HeJohn} in order to present an explicit web of connections relating
important mathematical structures in the study of $SU\left(2\right)$ Gaiotto theories. This
is the backbone of connections in Figure 2. Undertaking this task has allowed us to pinpoint the precise
manner in which dessins d'enfants arise in the context of these theories. Our conclusions are as follows:

\begin{itemize}

\item At a certain point in the Coulomb branch $\mathcal{U}_{g,n}$, the quadratic differential on $\mathcal{C}$ for the
Gaiotto theory in question is Strebel. At such a point, the horizontal trajectories
join to form a graph on $\mathcal{C}$ known as a \emph{ribbon graph} \cite{Mulase}. When the edges of this ribbon
graph are of equal length (found by fixing a particular point in $\mathcal{U}_{g,n} \times \mathbb{R}^n_+$), this graph can be interpreted as a clean dessin.
\item The ribbon graph, interpreted as a dessin, has a unique corresponding Belyi map
$\beta:\mathcal{C}\rightarrow\mathbb{P}^1$. This Belyi map relates
the Strebel differential on $\mathcal{C}$ at this point in $\mathcal{U}_{g,n} \times \mathbb{R}^n_+$ and a meromorphic quadratic differential on $\mathbb{P}^1$ by pullback.
\item By Belyi's theorem, the fact that this is possible for almost all Gaiotto theories 
means that almost all Gaiotto curves have the structure of algebraic curves defined over $\overline{\mathbb{Q}}$, at these particular points in $\mathcal{U}_{g,n} \times \mathbb{R}^n_+$.
\item Consideration as to the topology of the ribbon graphs yields a means of computing the essential features
of the quadratic differential in question:~it must have $n$ second order poles and $2n-4+4g$ zeroes. Possible
ribbon graph topologies for a Gaiotto theory with $\mathcal{C}$ having $n$ punctures and genus $g$ therefore
have $n$ faces and $2n-4+4g$ vertices.
\item This yields an efficient means of computing the explicit Strebel differentials, and hence Seiberg-Witten curves,
at these points in $\mathcal{U}_{g,n} \times \mathbb{R}^n_+$:~for the $\langle g, n \rangle$ Gaiotto theory in question, we compute all possible trivalent connected graphs with  $n$ faces and $2n-4+4g$ vertices, interpret as dessins, compute the associated Belyi maps, and substitute into (3.8).
\item The dessins in \cite{YMR} correspond to possible ribbon graphs of specific $SU\left(2\right)$ Gaiotto 
theories, which have been identified.
\item Ribbon graphs appear at points in the Coulomb branch where the triangulation (and hence BPS quiver) jumps.
\item In \cite{FirstCachazo, CachazoDessins}, it was found that the problem of finding Argyres-Douglas singularities
 for $U\left(N\right)$ $\mathcal{N}=2$ gauge theories can be mapped to the problem of finding points where an abstractly defined quadratic differential becomes Strebel; and therefore mapped to the problem of constructing dessins. We have found that there are difficulties in straightforwardly extending
 this story to the $SU\left(2\right)$ Gaiotto theories under consideration in this paper.
 
\end{itemize}

These conclusions establish the ``lower loop'' of connections in Figure 2,
as well as fleshing out many more details. The means of immediately writing down the functional form
of the quadratic differential $q$ given  topology $\left\langle g,n\right\rangle $
of the skeleton diagram is the ``upper arc'' of Figure 2.
Moreover, we have shown in \S4 that the method proposed in \cite{HeJohn} for writing down 
the Seiberg-Witten curve for such a theory by interpreting the skeleton diagram as a dessin must be modified in general.

There are many possible extensions of this work. Most notably, it would be a valuable task to understand better the physical significance of the points in the moduli space where ribbon graphs can be constructed (i.e.~the Strebel points), beyond the observation that these points lie on boundaries separating BPS domains of the Coulomb branch. Indeed, the authors are currently collaborating on a further paper investigating these Strebel points from the point of view of Liouville conformal field theories via the AGT conjecture \cite{AGT}; the hope is that such investigations will shed further light on the significance of these points, and the dessins that arise in these $\mathcal{N}=2$ theories.

In addition, it would be interesting to investigate whether
the Seiberg-Witten curves for these theories have any interesting factorisation properties at the points 
in the Coulomb branch at which the quadratic differential becomes Strebel and the edge lengths of the ribbon graph are fixed to be equal. Doing so would salvage some
connections and parallels with the work of \cite{FirstCachazo, CachazoDessins}. More generally, 
it would be an interesting and worthwhile task to carry out these investigations into Gaiotto theories of
higher rank; this is likely to be a fertile and fascinating
field for future research.

\section*{Acknowledgements}

This paper developed out of a fourth year undergraduate MPhys project undertaken by J.R.~at the University of Oxford in 2013, under the supervision of Y-H.H. and in correspondence with Diego Rodriguez-Gomez. We are particularly indebted to Diego for his work on earlier drafts of this paper, and for his invaluable insights and comments throughout. We are also grateful to Amihay Hanany and John McKay for helpful guidance, as well as to the anonymous reviewer for insightful comments. Y-H.H.~is indebted to the Science and Technology Facilities
Council, UK, for grant ST/J00037X/1; the Chinese Ministry of Education, for a Chang-Jiang
Chair Professorship at NanKai University; the city of Tian-Jin for a Qian-Ren Award; and Merton College, Oxford for their continued support. J.R.~is supported by an AHRC scholarship at the University of Oxford, and is also grateful
to Trinity College, Cambridge for a Studentship in Mathematics in 2013/14, and to Merton College, Oxford, for support.


\end{document}